\documentclass[traditabstract]{aa}

\usepackage{graphicx}
\pdfoutput=1
\usepackage{amsmath}
\usepackage{txfonts}
\usepackage[round]{natbib}
\newcommand{\Msun}{\mathrm{M}_{\odot}}
\newcommand{\Rsun}{\mathrm{R}_{\odot}}

\usepackage[usenames,dvipsnames]{color}

\begin{document}
   \title{Wind Roche-lobe overflow: \\Application to carbon-enhanced metal-poor stars}

   \author{C. Abate
          \inst{1}
          \and
          O. R. Pols \inst{1} 
          \and
          R. G. Izzard \inst{2}
          \and S. S. Mohamed \inst{3, 2} 
          \and S. E. de Mink \inst{4, 5}\thanks{Hubble fellow.}
          }

   \institute{Department of Astrophysics/IMAPP, Radboud University Nijmegen, P.O. Box 9010, 6500 GL Nijmegen, The Netherlands\\
              \email{[C.Abate; O.Pols]@astro.ru.nl}
         \and
             Argelander Institut f\"ur Astronomie, Auf dem H\"ugel 71, D-53121 Bonn, Germany\\
              \email{izzard@astro.uni-bonn.de}
         \and
             South African Astronomical Observatory, Observatory Road, Observatory 7925, South Africa\\
             \email{mohamed@saao.ac.za}
         \and
             Space Telescope Science Institute, 3700 San Martin Drive, Baltimore, MD, USA\\
             \email{demink@stsci.edu}
         \and
             Dep. of Physics and Astronomy, Johns Hopkins University, 3400 N. Charles Street, Baltimore, MD 21218, USA\\
             }

   \date{Received ...; accepted ...}
 
  \abstract
   {Carbon-enhanced metal-poor stars (CEMP) are observed as a substantial fraction of the very metal-poor stars in the Galactic halo. Most CEMP stars are also enriched in $s$-process elements and these are often found in binary systems. This suggests that the carbon enrichment is due to mass transfer in the past from an asymptotic giant branch (AGB) star on to a low-mass companion. Models of binary population synthesis are not able to reproduce the observed fraction of CEMP stars without invoking non-standard nucleosynthesis or a substantial change in the initial mass function. This is interpreted as evidence of missing physical ingredients in the models.
   Recent hydrodynamical simulations show that efficient wind mass transfer is possible in the case of the slow and dense winds typical of AGB stars through a mechanism called wind Roche-lobe overflow (WRLOF), which lies in between the canonical Bondi-Hoyle-Lyttleton (BHL) accretion and Roche-lobe overflow. WRLOF has an effect on the accretion efficiency of mass transfer and on the angular momentum lost by the binary system. The aim of this work is to understand the overall effect of WRLOF on the population of CEMP stars. To simulate populations of low-metallicity binaries we combined a synthetic nucleosynthesis model with a binary population synthesis code. In this code we implemented the WRLOF mechanism.
   We used the results of hydrodynamical simulations to model the effect of WRLOF on the accretion efficiency and we took the effect on the angular momentum loss into account by assuming a simple prescription. The combination of these two effects widens the range of systems that become CEMP stars towards longer initial orbital periods and lower mass secondary stars. As a consequence the number of CEMP stars predicted by our model increases by a factor $1.2-1.8$ compared to earlier results that consider the BHL prescription. Moreover, higher enrichments of carbon are produced and the final orbital period distribution is shifted towards shorter periods.}
 

   \maketitle

\section{Introduction}

The stellar population of the Galactic halo is characterised by low-mass stars with poor metal content. These stars are among the oldest that we observe, relics of the early stages of star formation in the Milky Way. The wide-field spectroscopic surveys HK \citep{BeersAJ92} and HES \citep{ChristliebAA01}, which are devoted to studying this population, reveal a surprisingly high frequency of carbon-enhanced metal-poor stars ($[\mathrm{C}/\mathrm{Fe}]\footnote{given two elements X and Y, their abundance ratio is [X/Y]$= \log_{10} (N_X/N_Y) - \log_{10} (N_{X\odot}/N_{Y\odot})$, where $N_{X,Y}$ refers to the number density of the elements X and Y and $\odot$ denotes the abundance in the Sun.} \ge +1.0$, CEMP stars hereinafter) among the very metal-poor stars (VMP, here indicating $[\mathrm{Fe}/\mathrm{H}]\lesssim-2.0$). The observed CEMP to VMP ratio is approximately 20\% (for example: $25\%$ \citealp{Marsteller2005}; $9\pm2\%$ \citealp{Frebel2006}; $21\pm2\%$ \citealp{Lucatello2006}), with the fraction of CEMP stars rapidly increasing for decreasing iron content and for increasing distance from 
the Galactic plane \citep{Carollo2012}.

CEMP stars are classified into different groups according to the presence of the heavy elements barium and europium, which are produced by slow and rapid neutron-capture processes ($s$-process and $r$-process), respectively. The largest group of CEMP stars are the $s$-process rich CEMP-$s$ stars, which display barium enhancements of [Ba/Fe]$>+0.5$ and account for at least 80\% of all CEMP stars \citep{Aoki2007}. Among these stars, some show enhancements of both $r$- and $s$-elements. A single case of a CEMP star highly enhanced only in $r$-process elements is also documented \citep{Sneden2003b}. Finally, one group of stars does not exhibit peculiar abundances of neutron-capture elements (e.g. \citealp{Aoki2002}). \cite{BeersARAA05} and \cite{Masseron10} provide detailed reviews of the CEMP subgroups.

Several hypotheses have been put forward to explain the overabundances observed in CEMP stars: ({\it a}) the level of carbon is primordial, or close to primordial, and was produced in the first generation of stars (e.g. \citealp{Mackey2003}, and \citealp{Cooke11}); ({\it b}) low-mass stars of extremely low metallicity might have undergone exotic mixing episodes that dredged up internally produced carbon to the surface \citep{Fujimoto-Ikeda-Iben00}; ({\it c}) a binary scenario in which in the past carbon-rich material from a thermally-pulsing asymptotic giant branch (TPAGB) primary star polluted the low-mass main-sequence secondary star. Today we only observe the secondary. The primary has become an unseen white dwarf. The binary scenario is currently considered the most likely formation mechanism for CEMP-s stars: in fact, the analysis performed by \cite{Lucatello05b} on a sample of these objects demonstrates that the fraction of CEMP-s stars with detected radial-velocity variations is consistent with the hypothesis of all being members of binary systems. The same binary mass transfer scenario is invoked to explain the properties of Ba and CH stars \citep{McClure90}.

A quantitative understanding of the origin of CEMP stars involves many branches of stellar astrophysics, some of which are still not well understood. The main uncertainties are related to ({\it i}) stellar evolution, particularly the nucleosynthesis during the AGB phase and internal mixing and diffusion processes in both stars of the binary system, ({\it ii}) the mass transfer process, ({\it iii}) the binary fraction of low-mass stars in the Halo and the distribution of orbital parameters, ({\it iv}) the initial mass function (IMF) at low metallicity. 

Several studies have considered population models in an attempt to reproduce the observed CEMP fraction of 9--25\%. \cite{Lucatello05a} and \cite{Komiya07} come to the conclusion that an IMF biased toward intermediate-mass stars is required to reproduce the fraction of CEMP/VMP stars measured in the Halo. However, large changes in the IMF are inconsistent with the small fraction of nitrogen-enhanced metal-poor stars observed in the Halo \citep{Pols2012}.
With a different approach, \citet[I09 hereinafter]{Izzard09} choose the solar neighbourhood IMF proposed by \citet[KTG93]{Kroupa93} in their population synthesis model and investigate the properties of metal-poor stars. I09 try to reproduce the observed fraction of CEMP/VMP\footnote{In I09 stars with [Fe/H] $\lesssim-2~$ are called {\it EMP} stars.} stars by only varying uncertain physical parameters related to nucleosynthesis, mass transfer and mixing processes.
For standard values of these physical parameters I09 find a CEMP/VMP fraction lower than the observed one by almost a factor 10. With a set of parameters that reduces the minimum core mass required for third dredge up and allows efficient third dredge up in stars of mass down to $0.8\,\Msun$ I09 find a CEMP/VMP fraction of approximately $9\%$, approaching the range of the observations. However, the observed distributions of carbon, nitrogen and heavy elements in CEMP stars are not well reproduced by the model.
This suggests that the above mentioned uncertainties need to be further investigated. In this paper we study the mass transfer process in some detail.

Recently, different indications have emerged suggesting that the efficiency of wind mass transfer in binary systems has so far been underestimated, at least in the case of slow and dense winds characteristic of AGB stars. Wind mass transfer plays an important role in the binary scenario for the formation of CEMP stars, and, more generally, is involved in several problems of astrophysical interest, e.g. the formation of CH and barium stars, the shaping of planetary nebulae, symbiotic stars, novae and the evolutionary path leading to the progenitors of Type Ia supernovae.
Therefore in this paper we push forward the analysis of I09 with a more accurate description of the wind mass transfer process and we investigate the effects of wind mass transfer on a population of binary stars, focusing in particular on CEMP stars.

In Sect. \ref{m-t} we briefly describe the context of wind mass transfer. In Sect. \ref{models} we discuss the main parameters of our binary population synthesis model and how wind mass transfer is implemented in our model. In Sect. \ref{results} and \ref{discussion} the results of our analysis are shown and discussed while Sect. \ref{conclusions} concludes.

\section{Wind mass transfer}
\label{m-t}

In the binary scenario, the primary star produces carbon and $s$-elements during the late stages of its evolution, when it undergoes thermal pulses. Dredge-up episodes bring these elements to the surface of the star, where they may be lost in a wind or by Roche-lobe overflow, polluting the main-sequence companion. The material accreted by the secondary star might in turn be diluted and burnt. When the donor is an AGB star, Roche-lobe overflow is typically unstable \citep{Paczynski1965} and will lead the binary system to a common envelope phase, with negligible accretion to the secondary (see \citealp{RickerTaam08} for more details on accretion in a common envelope). Therefore the wind mass transfer scenario plays a crucial role in CEMP formation.

\subsection{Limitations of the standard scenario}
The canonical description by \cite{BoHo} of the wind mass transfer mechanism is appropriate under the assumption that the wind velocity is much higher than the orbital velocity of the accreting star. This condition is not always fulfilled by AGB winds. In binary stars of periods around $10^4$ days the orbital velocity is about $10~\mathrm{km\,s}^{-1}$, whereas outflows from AGB stars are observed with wind velocities in the range $5-30$ $\mathrm{km\,s}^{-1}$. The mechanism that drives the wind is not fully understood. It is thought that in the outer atmosphere of the AGB star, where the temperature is low enough, dust grains form and are accelerated by radiation pressure, dragging the surrounding gas along due to collisional momentum transfer \citep{Hofner2009, Hofner2012}. 
Therefore the ratio between the orbital velocity and the wind velocity, which in turn determines the applicability of the Bondi-Hoyle-Lyttleton (BHL) prescription, depends on the separation and the mass of the two stars during the phase of mass transfer and on the chemical composition of the grains. 

\subsection{The Wind Roche-lobe overflow mechanism}
\label{WRLOF-mechanism}
In the past few years the possible existence of a relatively efficient mode of wind mass accretion in a binary system has emerged both from observations and hydrodynamical simulations. Mira AB \citep{Karovska05} and SS Leporis \citep{Blind2011} are two examples of detached binary systems that are undergoing very efficient mass transfer (about the efficiency of mass transfer in Mira AB there is some debate, see e.g. \citealp{Sokoloski2010}) although the AGB donor star is unambiguously filling only a fraction of its Roche lobe (approximately 10\% in the case of Mira,  80\% for SS Lep).
Moreover, recent hydrodynamical simulations suggest a new mode of mass transfer that lies somewhere in between Roche-lobe overflow and wind mass transfer. 
In this mode, called ``wind Roche-lobe overflow'' \citep[hereinafter WRLOF]{Shazrene07} or ``gravitational focusing'' \citep{deValBorro09}, the wind of the primary star is focused towards the orbital plane and in particular towards the secondary star. This has two major effects on the evolution of a binary system. The first effect is that the accretion rate predicted in the WRLOF regime is significantly higher than the BHL predictions. The second effect is that most of the material that is not accreted is lost from the vicinity of the outer Lagrangian points $L_2$ and $L_3$, with consequences for the angular momentum lost by the binary system that we further discuss in Sect. \ref{WRLOF-am}.

\begin{figure}[!]
\begin{center}
  \includegraphics[width=0.45\textwidth]{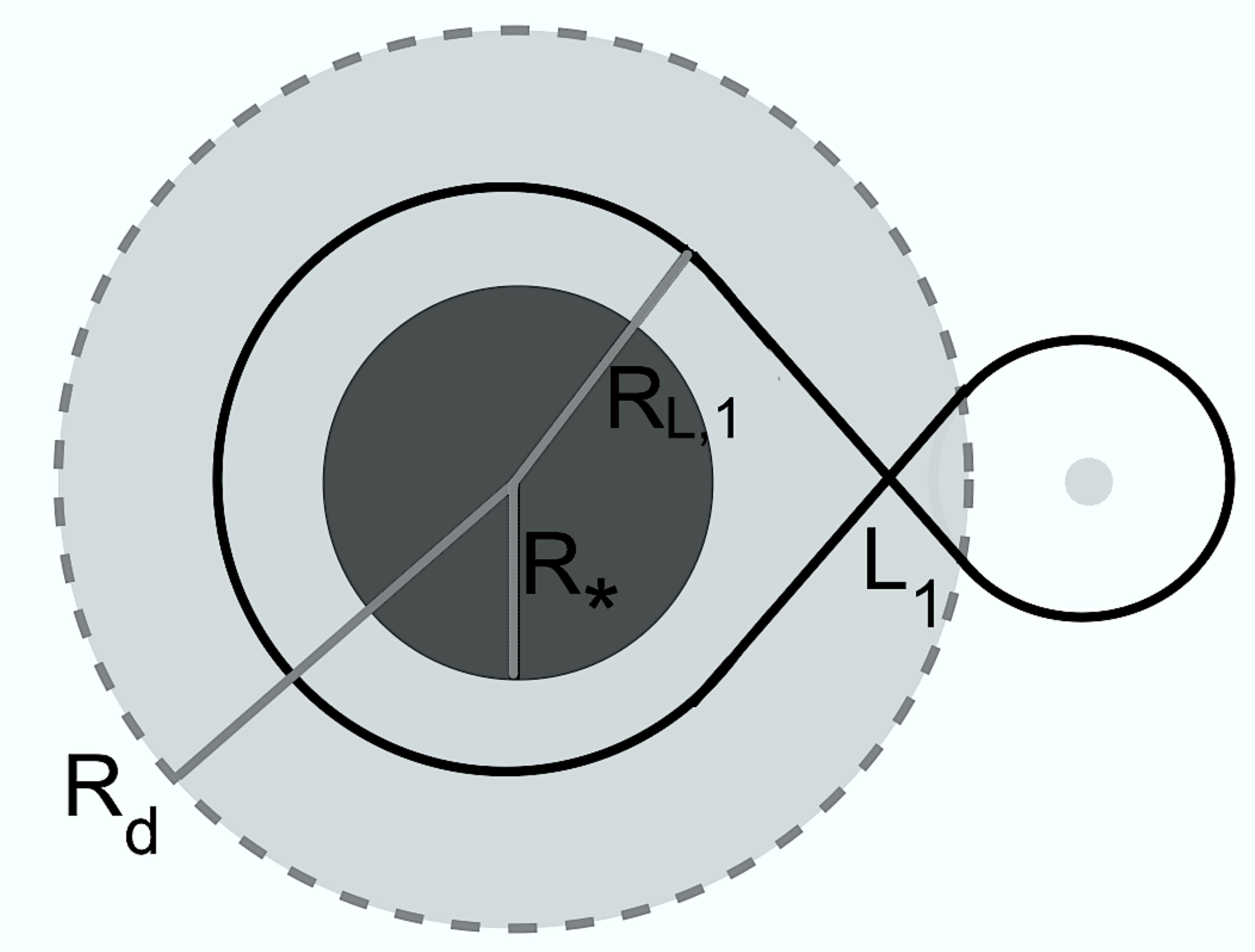}
\caption{Schematic picture of the WRLOF mechanism: $R_{\mathrm{d}}$, the wind acceleration radius, lies close to $R_{\mathrm{L,1}}$, the Roche-lobe radius. Inside the wind acceleration zone (shaded area) the wind is slow and can be efficiently accreted to the secondary through the first lagrangian point $L_1$ (sizes are not in scale).}
\label{fig:sketch}
\end{center}
\end{figure}

WRLOF can occur in systems where the wind is gravitationally confined to the Roche lobe of the primary star (the donor) and then falls into the potential well of the secondary through the inner Lagrangian point $L_1$ \citep{Shazrene07}. A good indicator of the conditions for the occurrence of WRLOF is the ratio $R_{\mathrm{d}}/R_{\mathrm{L,}1}$, where: 
\begin{itemize}
\item $R_{\mathrm{L,}1}$ is the Roche-lobe radius of the primary star. For a given mass ratio this quantity is proportional to the binary separation \citep{Kopal}.
\item $R_{\mathrm{d}}$ is the radius of the wind acceleration zone, namely the region where the wind is accelerated beyond the escape velocity. 
\end{itemize}
WRLOF can occur in systems where the wind acceleration radius is larger than, or is a significant fraction of, the Roche-lobe radius of the wind-losing star. This condition is schematically represented in Fig. \ref{fig:sketch}, where the wind acceleration zone is represented as a shaded area around the star of radius $R_*$.

In AGB stars we assume the acceleration of the wind is driven by radiation pressure on dust grains and therefore $R_{\mathrm{d}}$ coincides with the dust formation radius. In this case $R_{\mathrm{d}}$ scales with the stellar radius, and it also depends on the effective temperature of the AGB star, $T_{\mathrm{eff}}$, and on the chemical composition of the dust. \cite{Hofner2007} suggests the following approximate relation:
\begin{equation}
R_{\mathrm{d}} = \frac{1}{2}~R_*~\left(\frac{T_{\mathrm{eff}}}{T_{\mathrm{cond}}}\right)^{2.5}  \label{eq:dustradius}
\end{equation}
$T_{\mathrm{cond}}$ is the condensation temperature of the dust, a parameter that depends on the chemical composition of the compound, for example: $T_{\mathrm{cond}} = 1500$ K for carbon-rich dust (C/O$>1$) and $T_{\mathrm{cond}} = 1000$ K for oxygen-rich dust (C/O$<1$). During most of the AGB phase the dust formation radius is linearly proportional to the stellar radius and for carbon-rich dust $R_{\mathrm{d}}/R_*\approx 3$.

\cite{Shazrene10} uses a smoothed-particle hydrodynamics (SPH) code to simulate wind mass transfer in Mira-like circular binary systems with a $1\mathrm{M}_{\odot}$ primary in the AGB phase and a $0.6~\mathrm{M}_{\odot}$ secondary with separations between 5 and 60 AU. A similar approach is used by \cite{deValBorro09} but with $M_1=1.2~\mathrm{M}_{\odot},~M_2=0.6~\mathrm{M}_{\odot}$ and {\it a} varying up to 70 AU.
The evolution of each binary is followed in detail at least until the wind material has expanded to twice the size of the primary's Roche-lobe radius, and typically for an amount of time equal to many orbital periods. This amount of time is considered long enough to study the development of wind anisotropies, spiral shocks, equatorial flows, and to follow the fate of the lost material, distinguishing the fraction of mass that is accreted from the fraction that is lost or remains bound to the system but is not accreted. 

The wind is modelled by inserting particles in shells at the surface of the star, where the time interval between successive injections is regulated to produce a continuous outflow of material and to minimise discontinuities between the shells. 
Different types of wind are modelled: the one that best reproduces the properties of an AGB star like Mira considers a wind outflow with initial speed of 5.5 $\mathrm{km\,s}^{-1}$ at the stellar surface, and includes radial pulsations implemented using a piston approximation (e.g. \citealp{Freytag2008}) and radiation pressure on dust grains.
In every model the evolution in time of the positions and fluid properties is followed, predicting among other parameters the dust-formation radius, the ratios of the density, velocity, wind anisotropy at the equator relative to the poles, and the accretion efficiency.

Several parameters are likely to influence the mass transfer mechanism in Mira-type binaries, first of all the mass-loss rate, which is determined by the luminosity, temperature and pulsation period of the wind-losing star, although it is chosen to be approximately $10^{-6} \,\Msun\,$yr$^{-1}$ in the simulations. The model for the formation of the dust grains and the C/O ratio affects the velocity of the winds and therefore the intensity of WRLOF features. Moreover, stellar rotation and magnetic fields are likely to play an important role in shaping the outflow. 

In Table \ref{table:beta-vs-RdRRL} we list orbital periods, $R_{\mathrm{d}}/R_{\mathrm{L,}1}$ ratios and accretion efficiencies $\beta_{\mathrm{acc}}$, defined as the ratio between the mass accreted by the secondary star and the mass lost by donor star per unit of time. These values are the result of the simulations of \cite{Shazrene10} with the model of the AGB wind described earlier in this Section in which the evolution of every binary system is followed for at least one orbital period. Tests performed by \cite{Shazrene10} with different assumptions about the wind driving mechanism give an estimate of the uncertainty in $\beta_{\mathrm{acc}}$ which is within $50\%$ of each value (e.g. for $P=1.34\times10^5$ days $\beta_{\mathrm{acc}} = 0.10\pm0.05$).

The data in Table \ref{table:beta-vs-RdRRL} are shown as plus signs in Fig. \ref{fig:WRLOF-aq} together with a proposed model (solid line) which is further discussed in Sect. \ref{WRLOF-model}. High $R_{\mathrm{d}}/R_{\mathrm{L,}1}$ values correspond to close systems whereas low values correspond to wide systems because $R_{\mathrm{d}}/R_{\mathrm{L,}1}\propto R_*/a$. 
\begin{table}
\caption{Initial period $P$, ratio between the dust formation radius and the Roche-lobe radius, $R_{\mathrm{d}}/R_{\mathrm{L,}1}$, and accretion efficiency $\beta_{\mathrm{acc}}$ for binary systems with a $1.0\,\Msun$ AGB primary and a $0.6\,\Msun$ main-sequence companion (from \citealp{Shazrene10}). Because $R_{\mathrm{L,}1}  \propto a$ (and $R_{\mathrm{d}}$ does not depend on $a$), high values of the $R_{\mathrm{d}}/R_{\mathrm{L,}1}$ ratio correspond to close systems and low values correspond to wide systems.}          
\label{table:beta-vs-RdRRL}   
\centering        
\begin{tabular}{c c c}
\hline
\hline
$P/$days & $R_{\mathrm{d}}/R_{\mathrm{L,}1}$ & $\beta_{\mathrm{acc}}$ \\ 
\hline
$1.34\times10^5$&0.40 &0.10\\
$5.98\times10^4$&0.67 &0.24 \\
$2.58\times10^4$&1.18 &0.45 \\
$9.12\times10^3$&2.4 &0.35 \\
$3.21\times10^3$&2.8 &0.10 \\
\hline 
\end{tabular}
\end{table}
From wide to close systems (i.e. from left to right in Fig. \ref{fig:WRLOF-aq}) the accretion efficiency $\beta_{\mathrm{acc}}$ initially increases, then reaches a maximum and finally decreases again with $R_{\mathrm{d}}/R_{\mathrm{L,}1}$.
This behaviour is explained as follows: for a given $M_1$ and $M_2$, in wide systems the wind acceleration zone is smaller than the Roche lobe, therefore the wind is fast and only a small fraction of it is accreted, as also predicted by the BHL prescription. Moving to smaller separations (i.e. greater $R_{\mathrm{d}}/R_{\mathrm{L,}1} $) the dust formation radius becomes a progressively more significant fraction of the Roche-lobe radius, the wind is increasingly confined in the primary's Roche lobe and therefore the accretion efficiency grows. With smaller $a$ the accretion efficiency decreases because in close systems a large fraction of the wind escapes through $L_2$ and $L_3$ and is not accreted to the secondary star. We refer to \cite{Shazrene07} and to \cite{Shazrene10} for further details about the WRLOF mechanism.
For the purposes of our work we apply the results of the above-mentioned hydrodynamical simulations to our binary population synthesis code, as described later in Sect. \ref{WRLOF-model}, and we evaluate how this mode of mass transfer affects the predictions for CEMP stars in a population of binary systems.

\section{Models}
\label{models}
In order to perform this study we make use of the binary population synthesis code described by \cite{Izzard04,Izzard06,Izzard09}. Our model follows binary evolution according to the rapid binary stellar evolution prescription of \cite{Hurley02} and the algorithms for AGB evolution and nucleosynthesis of \cite{Izzard04,Izzard06}. For a complete discussion of our model, its characteristics, parameters and uncertainties we refer to I09.
In this section we briefly summarise the most important characteristics of our model (Sect. \ref{popsyn}), the updates that we introduce compared to the work by I09 (Sect. \ref{updates}) and the way we implement WRLOF (Sect. \ref{WRLOF-model}).

\subsection{Population synthesis}
\label{popsyn}
Our population synthesis simulations are based on a grid of $N^3$ binary evolution models distributed uniformly in $\ln M_1 - \ln M_2 - \ln a $ parameter space, where $M_1$ and $M_2$ are the initial masses of the primary and of the secondary star respectively, $a$ is the initial separation of the system and we take $N = 128$. The initial metallicity of our model sets is $Z = 10^{-4}$, or equivalently [Fe/H]$=-2.3$. We consider circular orbits, therefore the eccentricity, $e$, is always zero.

We count the stars of a particular type, for example CEMP stars, according to the sum
\begin{eqnarray}
n_{\mathrm{type}} = S \sum_{M_{1,\mathrm{min}}}^{M_{1,\mathrm{max}}}\sum_{M_{2,\mathrm{min}}}^{M_{2,\mathrm{max}}}\sum_{a_{\mathrm{min}}}^{a_{\mathrm{max}}}\sum_{t_{\mathrm{min}}}^{t_{\mathrm{max}}} \delta_{\mathrm{type}}~ \Psi_{M_1,M_2,a}~ \delta M_1~\delta M_2~ \delta a~ \delta t~, \label{eq:nstars}
\end{eqnarray}
where, as in the paper by I09:
\begin{itemize}
\item the size of a cell in the parameter space is $\delta M_1 \cdot \delta M_2 \cdot \delta a$; the timestep is $\delta t$.
\item $S$ is the star formation rate which is assumed to be constant;
\item $\delta_{\mathrm{type}}$ is equal to 1 when the star is of the required type and is zero otherwise. Stars are selected from our model population according to their age, surface gravity and surface abundances, as follows: {\bf VMP} stars are older than 10 Gyr and with surface gravity $\log_{10} (g/\mathrm{cm}\,\mathrm{s}^{-2}) \le 4.0\,$; {\bf CEMP} stars are VMP stars characterised by a surface abundance of carbon [C/Fe]$\ge 1.0\,$; {\bf CEMP-s} stars are CEMP stars also enriched in barium, [Ba/Fe]$\ge0.5$.
\item $M_1$ and $M_2$ vary respectively in the ranges [$0.7,8.0$] $\Msun$ and [$0.1,0.9$] $\Msun$. Initially $M_2 \le M_1$ by definition.
\item $a$ varies between 3 and $10^5$ $\Rsun$. To be able to compare our work with the results of I09 we assume that all stars are formed in binary systems with this range of separations. In reality some VMP stars are single or in wider orbits and we can take these stars into account by reducing the binary fraction.
\item $t$ varies in the range [$10,13.7$] Gyr, the approximate ages of the Halo and the Universe.
\item $\Psi$ is the inital distribution of $M_1$, $M_2$, and $a$. We assume that $\Psi$ is separable,
\begin{equation}
\Psi = \Psi(M_1, M_2, a) = \psi(M_1)~\phi(M_2)~\chi(a)~,
\end{equation}
where the primary mass distribution $\psi(M_1)$ is the intial mass function by KTG93, the secondary mass distribution $\phi(M_2)$ is flat in $q = M_2/M_1$ (any mass ratio $\boldsymbol{0 < q \le 1}$ is equally likely), the separation distribution $ \chi(a)$ is flat in $\ln a$ (i.e. $\chi(a) \propto 1/a$).
\end{itemize}

Our model sets assume efficient thermohaline mixing: the accreted material mixes instantaneously with the stellar envelope. This approximation is reasonable in many cases, as suggested by the calculations of \cite{Stancliffe07}, even though more recent studies (e.g. \citealp{Stancliffe08}) show that the situation is more complicated and other processes such as gravitational settling in some cases reduce the effect of thermohaline mixing.

\subsection{Parameter choices and updates}
\label{updates}
The nucleosynthesis algorithm which follows the evolution of the star through the first, second and third dredge ups, modifying the surface abundances as required, is mostly based on the work by \cite{Karakas02} and \cite{Karakas07}. A prescription for hot-bottom burning is also included. Abundances are scaled according to the solar values by \cite{AG89}. Wind mass-loss rates are parameterised according to the \citet{Reimers75} formula multiplied by a factor $\eta=0.5$ on the first giant branch and \cite{VW93} on the AGB. The $^{13}\mathrm{C}$ pocket efficiency $\xi_{13}$, as defined by \cite{Izzard06}, is set to 1 by default. The common envelope evolution is treated according to the prescription of \cite{Hurley02} with $\alpha_\mathrm{CE}=1$, and we do not include accretion during the common envelope phase. 
Our default model set A uses the same input physics as the model set A described in Sect. 2.1.5 and 4.1 of the paper by I09. We list here the updates we introduce.
\begin{itemize}
\item {\it Wind velocity.} To calculate the mass accretion rate, I09 assume the wind velocity is a fraction of the escape velocity from the mass-losing star. However, the escape velocity decreases as the radius of the star and its mass-loss rate increase, whereas the wind velocities observed in TPAGB stars show an opposite trend. Therefore to calculate the mass accretion rate when the primary is a TPAGB star we take the wind velocity defined in Eq. (3) of the article by \cite{VW93}, which was derived by interpolating the observations. We choose a minimum value of $v_{\mathrm{w}} = 5  \mathrm{~km\,s^{-1}}$, which is consistent with the observations (e.g. \citealt{KnappMorris}). As \cite{VW93} we introduce a maximum value of $v_{\mathrm{w}} = 15  \mathrm{~km\,s^{-1}}$.\newline
\item {\it Angular momentum loss.} The algorithm used by I09 to calculate the change to the orbital angular momentum is based on Eq. (21) of the article by \cite{Hurley02}.
This equation only applies if the mass lost by the wind is much greater than the mass accreted by the secondary star. In most of our model sets we calculate the change to the orbital angular momentum assuming a spherically symmetric wind,
\begin{equation}
\dot{J} = \left[\left(\dot{M}_{1\mathrm{W}} - \dot{M}_{2\mathrm{A}} \right) M_2^2 +
\left(\dot{M}_{2\mathrm{W}} - \dot{M}_{1\mathrm{A}} \right) M_1^2
 \right] \frac{a^2 \Omega_{\mathrm{orb}}}{(M_1+M_2)^2}
~,~\label{eq:jeans}
\end{equation} 
where $\Omega_\mathrm{orb} = 2\pi/P$ is the orbital angular frequency and $\dot{M}_{i\mathrm{W}}$ and $\dot{M}_{i\mathrm{A}}$ are, respectively, the wind mass-loss rate and accretion rate of star $i$.
When the accretion efficiency is low, as in the BHL prescription, Eq. (21) of \cite{Hurley02} approaches Eq. (\ref{eq:jeans}). Therefore we do not expect the CEMP to VMP ratio to change significantly after this modification of the model. In fact, the fraction of CEMP to VMP stars that we compute with the two equations is the same: 2.30\% and 2.31\% (using respectively Eq. (21) of \citealp{Hurley02}, and Eq. (\ref{eq:jeans}), see Table \ref{tab:models}).
The model sets $\mathrm{A}'-\mathrm{C}_{\mathrm{q}}$ and $\mathrm{G}-\mathrm{J}_{\mathrm{q}}$ in Table \ref{tab:models} include Eq. (\ref{eq:jeans}) and we will refer to these as {\it ssw--}model sets. \\
\end{itemize}

\begin{table*}
\caption{Physical parameters in our binary population models and the derived CEMP/VMP ratio. Model set A is the same as in the article by I09. All models use the wind velocity prescription of \cite{VW93} except model sets A and A$'$ where $v_{\mathrm{w}}\propto v_{\mathrm{esc}}$. The errors convey only Poisson statistics and could be reduced by increasing the resolution $N$ of our grid of models. A finer grid is not necessary because the observed CEMP/VMP ratio varies between $9\%$ and $25\%$, a much wider range than the errors in our models.}
\label{tab:models}   
\centering        
\begin{tabular}{c c c c c}
\hline
\hline
Model set  & Ang. mom. loss & $ \beta_{\mathrm{acc}}$ & Third dredge up & CEMP/VMP (\%) \\ 
\hline
$A$ & as \cite{Hurley02} & BHL & standard & $2.30 \pm 0.04 $ \\
A$'$ & ssw -- Eq. (\ref{eq:jeans}) & BHL & standard & $2.31 \pm 0.04$ \\
\hline
B$_{~}$ & ssw -- Eq. (\ref{eq:jeans}) & BHL & standard & $2.22 \pm 0.04 $\\
C$_{~}$ & ssw -- Eq. (\ref{eq:jeans}) & WRLOF & standard & $2.90 \pm 0.04 $ \\
$\mathrm{C}_\mathrm{q}$ & ssw -- Eq. (\ref{eq:jeans}) & WRLOF $q$-dep & standard & $2.63 \pm 0.04 $ \\
D$_{~}$ & $\gamma2$ -- Eq. (\ref{eq:gamma2}) & BHL & standard  & $3.12 \pm 0.04 $ \\
E$_{~}$ & $\gamma2$ -- Eq. (\ref{eq:gamma2}) & WRLOF & standard  & $4.06 \pm 0.03 $ \\
$\mathrm{E}_{\mathrm{q}}$ & $\gamma2$ -- Eq. (\ref{eq:gamma2})& WRLOF $q$-dep & standard  & $3.85 \pm 0.04 $ \\
\hline
G$_{~}$ & ssw -- Eq. (\ref{eq:jeans}) & BHL & enhanced & $9.00 \pm 0.12 $ \\
J$_{~}$ & ssw -- Eq. (\ref{eq:jeans}) & WRLOF & enhanced & $11.15 \pm 0.13 $ \\
$\mathrm{J}_{\mathrm{q}}$ & ssw -- Eq. (\ref{eq:jeans}) & WRLOF $q$-dep & enhanced & $11.91 \pm 0.14 $ \\
I$_{~}$ & $\gamma2$ -- Eq. (\ref{eq:gamma2}) & BHL & enhanced & $10.27 \pm 0.13 $ \\
K$_{~}$ & $\gamma2$ -- Eq. (\ref{eq:gamma2}) & WRLOF & enhanced & $12.96 \pm 0.14 $ \\
$\mathrm{K}_{\mathrm{q}}$ & $\gamma2$ -- Eq. (\ref{eq:gamma2}) & WRLOF $q$-dep & enhanced & $13.66 \pm 0.14 $ \\
\hline 
\hline 
\end{tabular}
\end{table*}

\subsection{A proposed model for Wind Roche-lobe Overflow}
\label{WRLOF-model}
In the WRLOF mode of mass transfer the wind material is ejected in a highly aspherical geometry. This has two major effects, on the accretion efficiency of the mass transfer and on the angular momentum lost by the binary systems. We discuss our prescription to consider these two effects in Sect. \ref{WRLOF-beta} and Sect. \ref{WRLOF-am} respectively.

\subsubsection{The WRLOF accretion efficiency}
\label{WRLOF-beta}
\begin{figure}
   \centering
   \includegraphics[width=0.4\textwidth]{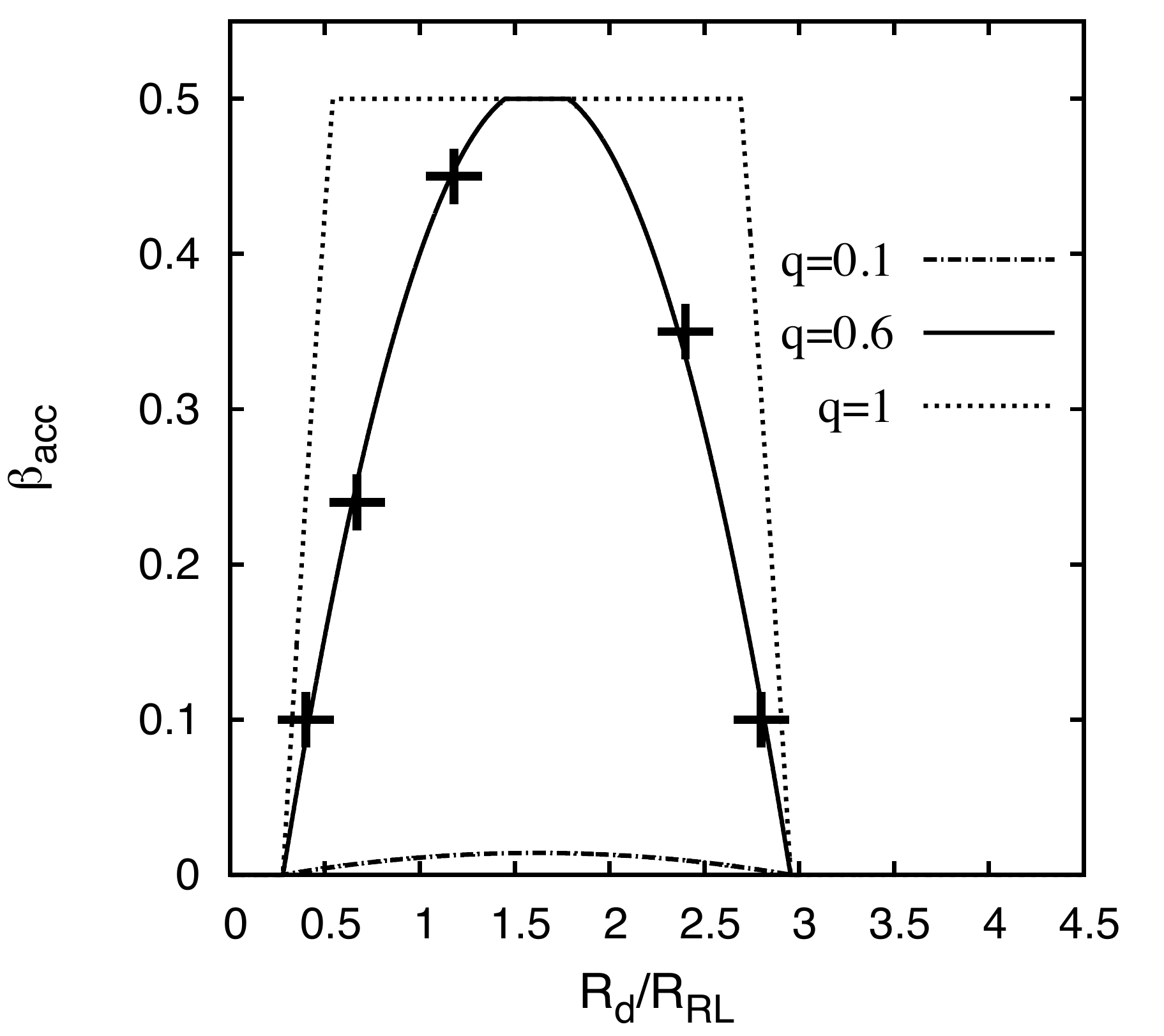}
      \caption{Models of the wind accretion efficiency $\beta_{\mathrm{acc}}$ for different values of $q$ calculated with Eq. (\ref{eq:WRLOF-aq}). The solid line is for $q=0.6$, the mass ratio used by \cite{Shazrene10}. The dot-dashed and dotted lines show the resulting efficiency for $q=0.1$ and $q=1$, respectively. Plus signs are the results of hydrodynamical simulations by \cite{Shazrene10} listed in Table~\ref{table:beta-vs-RdRRL}.}
         \label{fig:WRLOF-aq}
   \end{figure}

To use the results of the hydrodynamical simulations in our model, we fitted the data of Table \ref{table:beta-vs-RdRRL} with a parabolic function,
\begin{equation}
\beta_{\mathrm{acc}} = c_1~ x^2 + c_2~x + c_3~~,
\label{eq:WRLOF}
\end{equation}
where $x = R_{\mathrm{d}}/R_{\mathrm{L,}1}$, $c_1 = -0.284\, (\pm 0.018)$, $c_2 = 0.918\, (\pm 0.057)$ and $c_3 = -0.234\, (\pm 0.034)$. The errors result from the non-linear least-squares algorithm used for the fit and do not convey the uncertainty associated to the simulations.
Since in none of the models calculated by \cite{Shazrene10} the accretion efficiency is higher than $50\%$ we impose a maximum value $\beta_{\mathrm{acc,max}} = 0.5$ on $\beta_{\mathrm{acc}}$.
The result of the best fit is shown in Fig. \ref{fig:WRLOF-aq} as a solid line. 

Since Eq. (\ref{eq:WRLOF}) can assume zero or negative values in the range of possible values of $R_{\mathrm{d}}/R_{\mathrm{L,}1}$, we always choose the maximum accretion rate between WRLOF and BHL. This implies that for large separations, i.e. $R_{\mathrm{d}}/R_{\mathrm{L,}1}$ smaller than about 0.4, we calculate the accretion rate with the formulation of the BHL prescription given by \cite{BoffinJorissen},
\begin{equation}
\beta_{\mathrm{BHL}} = \frac{\alpha}{2\sqrt{1-e^2}} \cdot \left(\frac{GM_2}{a~v_\mathrm{w}^2}\right)^2~\left[1 + \left(\frac{v_{\mathrm{orb}}}{v_{\mathrm{w}}}\right)^2\right]^{-\frac{3}{2}} \label{eq:BoffinJorissen} ~,
\end{equation}
where $\alpha=1.5$ is a constant, $v_{\mathrm{orb}}$ is the relative orbital velocity, $G$ is the gravitational constant and the other quantities have the meaning already described. This is consistent with \cite{Shazrene10}, in which at large separations the same accretion rates are obtained as BHL.

\subsubsection{An example of accretion efficiency in the WRLOF regime}
\begin{figure}
   \centering
   \includegraphics[width=0.44\textwidth]{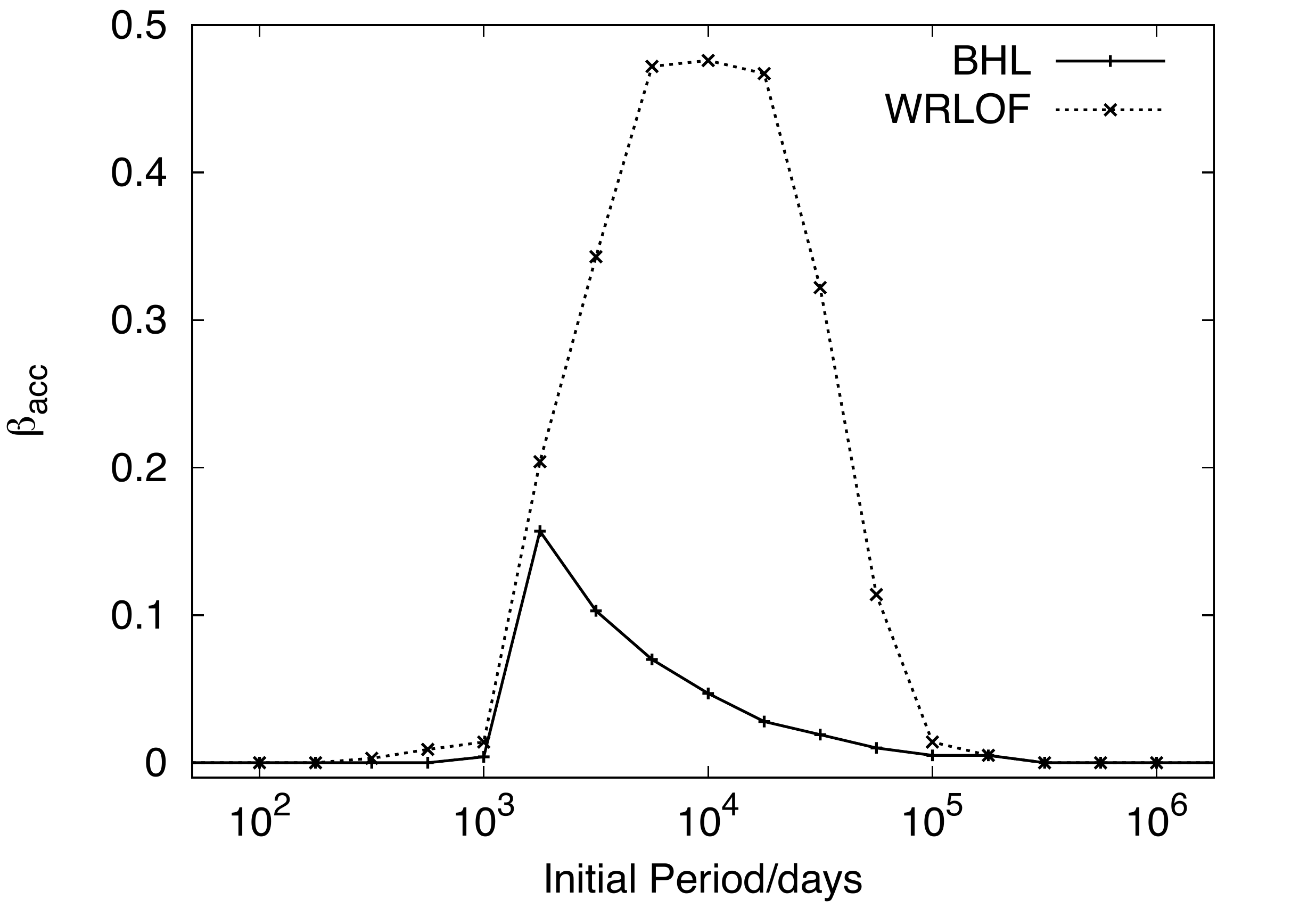}
      \caption{Efficiency of wind mass transfer $\beta_{\mathrm{acc}}$ as a function of the initial period for a binary system with $M_1=1.0\,\Msun$ and $M_2=0.6\,\Msun$ calculated with BHL (solid line) and WRLOF (dotted line).}
         \label{fig:WRLOF-effic}
   \end{figure}

To illustrate the effects of WRLOF on the wind mass transfer process we compute the accretion efficiency in binary systems with different initial periods and fixed initial masses, $M_1= 1.0\,\Msun$ and $M_2=0.6$ $\Msun$ as in the simulations of \cite{Shazrene10}. In Fig. \ref{fig:WRLOF-effic} we plot the accretion efficiency calculated with the BHL (solid line) and the WRLOF (dotted line) models as a function of the initial period. The maximum accretion efficiency in the case of WRLOF is almost 50\% and this high mass transfer efficiency is possible over a wide range of periods. A $1\,\Msun$ star loses approximately $0.2\, \Msun$ during the AGB phase, therefore its companion can reach $0.7\,\Msun$, close to the minimum mass for a metal-poor star to evolve off the main sequence in a Hubble time. The maximum accretion efficiency reached in the BHL model is roughly 15\% for periods around 1800 days, which means the companion accretes at most $0.03\,\Msun$ and typically much less.

\subsubsection{Mass-ratio dependence of the WRLOF efficiency}
Eq. (\ref{eq:WRLOF}) implicitly assumes that the accretion efficiency $\beta_{\mathrm{acc}}$ is independent of the mass ratio $q= M_2/M_1$. No conclusion can be drawn about any dependence on $q$ based on the results of \cite{Shazrene10} which are calculated for a fixed mass ratio $q=0.6$. Therefore, additional assumptions are required to introduce a relation between $\beta_{\mathrm{acc}}$ and $q$. In Eq. (\ref{eq:BoffinJorissen}) the dependence on $q$ for a fixed $M_1$ can be expressed as,
\begin{equation}
\beta_{\mathrm{BHL}} =  \frac{\alpha}{2\sqrt{1-e^2}} \cdot q^2 \left(\frac{GM_1}{a~v_\mathrm{w}^2}\right)^2~\left[1 + \underbrace{(1+q)\frac{G M_1}{a~v_{\mathrm{w}}^2}}_T\right]^{-\frac{3}{2}} \label{eq:BoffinJorissen2} ~.
\end{equation}
Hence the accretion efficiency $\beta_{\mathrm{BHL}}$ scales with a factor $q^2$ and there is an additional dependence related to the term $T$ in the square brackets. If term $T$ dominates, i.e. for small separations when $v_{\mathrm{orb}}>v_{\mathrm{w}}$, Eq. (\ref{eq:BoffinJorissen2}) becomes:
\begin{eqnarray}
\beta_{\mathrm{BHL}} \approx \frac{\alpha}{2\sqrt{1-e^2}} \cdot \frac{q^2}{(1+q)^{3/2}} \cdot \left(\frac{GM_1}{a~v_\mathrm{w}^2}\right)^{1/2} ~~. 
\end{eqnarray}
Vice versa when $T$ is not important, e.g. for large separations, then $\beta_{\mathrm{BHL}}$ scales simply as the mass ratio squared, $q^2$. For simplicity we choose to consider only the latter dependence in the model of WRLOF, keeping in mind that large separations correspond to small orbital velocities and therefore to the case when $v_{\mathrm{orb}}\lesssim v_{\mathrm{w}}$, the assumption under which the BHL prescription is appropriate.

Assuming this dependence on $q$ of the mass-accretion efficiency in the WRLOF we introduce the following expression:
\begin{equation}
\beta_{\mathrm{acc}} = \mathrm{min}\left\{\frac{25}{9} q^2 \left[c_1~ x^2 + c_2~x + c_3\right], ~\beta_{\mathrm{acc,max}} \right\}\label{eq:WRLOF-aq}
\end{equation}
where $x$, $c_1$, $c_2$, $c_3$, $\beta_{\mathrm{acc,max}}$ are as in Eq. (\ref{eq:WRLOF}) and the factor 25/9 arises because Eq. (\ref{eq:WRLOF-aq}) should coincide with Eq. (\ref{eq:WRLOF}) for $q=0.6$. 
In Fig. \ref{fig:WRLOF-aq} we show our models for different values of $q$. 

We point out that the term $T$ is not negligible in general. Therefore we expect the real dependence of $\beta_{\mathrm{acc}}$ on the mass ratio to be weaker than $q^2$. We consider Eqs. (\ref{eq:WRLOF}) and (\ref{eq:WRLOF-aq}) to bracket the dependence on $q$ of the accretion efficiency.

The calculations of \cite{Shazrene10} are for AGB primaries with oxygen-rich winds and dust, i.e. $\mathrm{C/O}<1$. Therefore the value of the parameters $c_1$, $c_2$ and $c_3$ of the fit could be different in the case of carbon-rich winds ($\mathrm{C/O}>1$), which is the relevant situation for CEMP stars. However, at the moment we cannot take this into account because there are no hydrodynamical simulations of binary systems with a carbon-rich primary star.

\subsubsection{Angular momentum loss in the WRLOF regime}
\label{WRLOF-am}
The hydrodynamical simulations performed by \cite{Shazrene10} show that the density of the matter lost in the wind is strongly enhanced towards the orbital plane. This enhancement is clear also in wider systems where the accretion efficiency is consistent with the BHL predictions. Because the geometry of the wind is not spherically symmetric, a prescription alternative to Eq. (\ref{eq:jeans}) is required to calculate the amount of angular momentum carried away by the wind. 

Let us assume that the material lost from the binary system carries away a multiple $\gamma$ of the average specific orbital angular momentum,
\begin{equation}
\dot{J} = \gamma \times \frac{J_{\mathrm{orb}}}{M_1+M_2} \left(\dot{M_1} + \dot{M_2}\right)~~.\label{eq:gamma2}
\end{equation}
If the matter is lost from the vicinity of $L_2$ and $L_3$ one expects efficient loss of angular momentum with $\gamma>1$.
However, the exact amount of angular momentum lost in a WRLOF situation is not well constrained by the hydrodynamical models of \cite{Shazrene10}. \cite{Jahanara2005} have studied angular momentum loss associated with wind mass transfer, also using hydrodynamical simulations but with different assumptions regarding the wind mechanism. In the case of a `radiative wind' -- which corresponds best to the case of an AGB wind -- with a velocity much lower than the orbital velocity, they find that matter leaving the binary system has an average specific angular momentum of approximately $0.6\times a^2\Omega_\mathrm{orb}$. This corresponds to a value $\gamma\approx$ 2.4--3.2 for mass ratios in the range 1/3--3.
In their study of the formation mechanism of barium stars \cite{Izzard10} use Eq. (\ref{eq:gamma2}) with $\gamma=2$ to reproduce the periods and eccentricities of the barium stars.
We use the same prescription in some of our models, i.e.\ in model sets $\mathrm{D}-\mathrm{E}_{\mathrm{q}}$, I, K, $\mathrm{K}_{\mathrm{q}}$ in Table \ref{tab:models} and we will refer to these as {$\it \gamma2-$}model sets. In these models we apply efficient angular momentum loss with $\gamma=2$ to all binary systems because the geometry of the wind is modified also in systems that are not in the regime of enhanced accretion.

We note that angular momentum loss from a spherically symmetric wind from the primary star, Eq.~(\ref{eq:jeans}) with $\dot{M}_{2\mathrm{W}}=0$, can be written in the form of Eq.~(\ref{eq:gamma2}) with $\gamma=M_2/M_1$.
Compared to this {\it ssw} situation, our {$\it \gamma2$} models with $\gamma=2$ correspond to a mode in which the binary system loses more orbital angular momentum and therefore shrinks more.

\section{Results}
\label{results}
We start our analysis with model set A which has the same input physics as the default model A in the article by I09 and predicts the same CEMP to VMP ratio of 2.30\%.
We successively modify model set A with the changes mentioned in Sect. \ref{updates} and \ref{WRLOF-model} and we study the effect of each of these updates on the overall population of CEMP stars. In Table \ref{tab:models} our different model sets are listed with the corresponding CEMP to VMP ratio.

\subsection{Model set B: wind velocity}
The wind-velocity relation of \cite{VW93} implies that $v_{\mathrm{w}}$ increases with time, reaches a maximum of 15 $\mathrm{km\,s}^{-1}$ in the regime of superwind and then remains constant until the end of the AGB phase. On the other hand, in model set A the wind velocity is directly proportional to the escape velocity, $v_{\mathrm{esc}}$, which decreases in time. 
Even though the dependence of the wind accretion efficiency on $v_{\mathrm{w}}$ is strong (see Eq. \ref{eq:BoffinJorissen}), in the regime of strong mass loss the actual difference between the two calculated values of $v_{\mathrm{w}}$ is small, within a factor 2. Therefore we find that in model set B the CEMP to VMP ratio is only slightly reduced to 2.22\%.

\subsection{Model set D: efficient angular momentum loss}
   \begin{figure}
   \centering
   \includegraphics[width=0.44\textwidth]{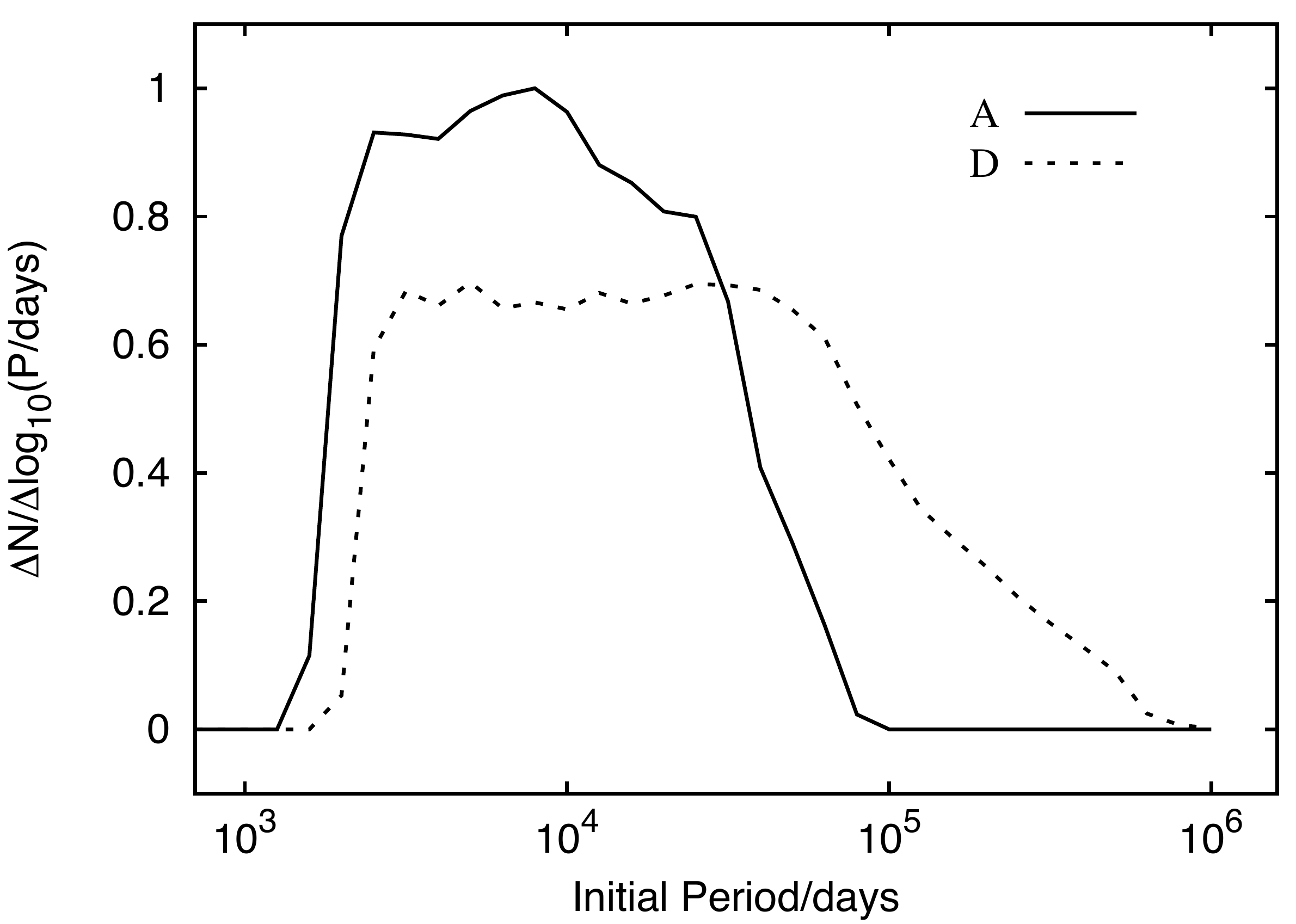}
   \includegraphics[width=0.44\textwidth]{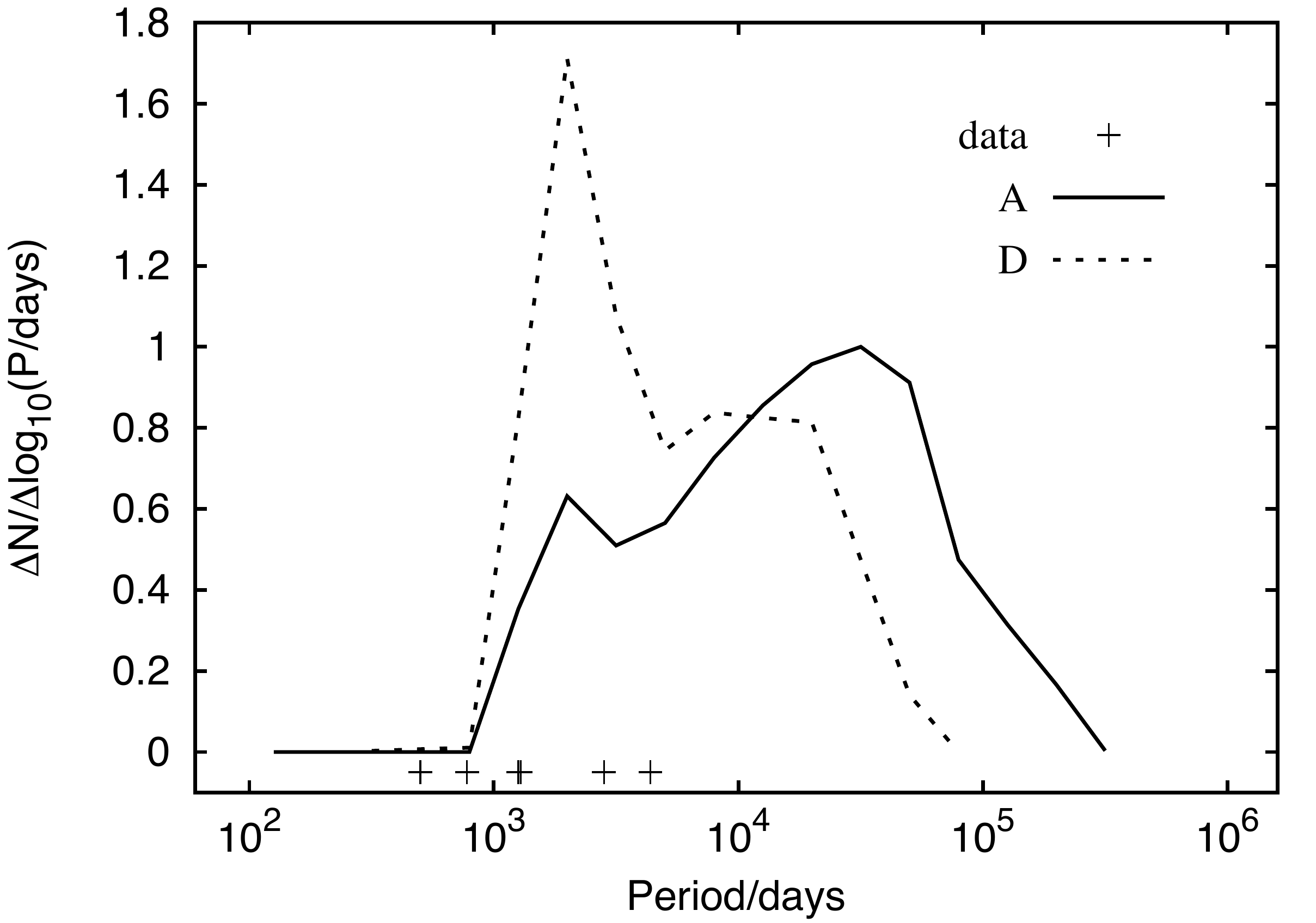}
      \caption{Initial ({\it top panel}) and final ({\it bottom panel}) period distributions of the CEMP population. In model set D (dotted line) angular momentum losses are computed with Eq. (\ref{eq:gamma2}) and $\gamma=2$. Solid line is the standard model set A. Plus signs are the periods of observed stars. The bins are equally spaced and the width in $\log_{10} (P/$days$)$ of each bin is $0.1$ (top panel) and $0.2$ (bottom panel). In this and in the plots that follow the $y$-axis indicates the expected number of CEMP stars in each bin and the plots are normalised such that the area under the graph is the same for each model and model A peaks at 1.}
         \label{fig:period-y2}
   \end{figure}
The hydrodynamical simulations of \cite{Shazrene10} show that the geometry of the wind ejected by the primary star is modified by the gravitational field of the companion in a wide range of systems, including those that are not in the regime of enhanced accretion efficiency. It is therefore instructive to study the effect of efficient angular momentum loss on a population of CEMP stars independently. In model set D the variations in the orbital angular momentum are computed using $\gamma=2$ in Eq. (\ref{eq:gamma2}), while the accretion rate is computed with the BHL prescription Eq. (\ref{eq:BoffinJorissen}).
Model set D predicts a CEMP to VMP ratio of $3.12\%$ and compared to model sets A and A$'$ produces notable differences in the distribution of initial and final period of the CEMP population.
The initial period range of systems that lead to a CEMP star is broadened and shifted towards wider separations (see Fig. \ref{fig:period-y2}, top panel). Systems shrink more due to wind mass loss and therefore wide binaries become close enough to efficiently transfer material. On the other hand, initially close binaries tend to undergo a common envelope phase, in some cases merge, and do not become CEMP stars. Analogously the final period distribution is shifted towards relatively close systems and the distribution peaks at a period more than one order of magnitude shorter than in set A, as shown in the bottom panel of Fig. \ref{fig:period-y2}. 

A direct comparison of our population with a measured period distribution is not possible: the number of CEMP stars with known period is low, despite the fact that many of them are known to be binaries, because long periods are difficult to measure. Since the average time for which surveys have been ongoing is about ten years, this is approximately the maximum period that we can presently measure.
In the observational SAGA database only six CEMP stars have a measured period (see Table 4 in the article by I09), and they are displayed in Fig. \ref{fig:period-y2} as plus signs (two systems have almost the same period, $1.26\times10^3$ and $1.29\times10^3$ respectively, and therefore are indistinguishable in the plot). The two shortest-period CEMP binaries are not reproduced by any of our models.
The characteristics of the distributions of other parameters, e.g. $M_1$, $M_2$, [C/Fe], are not significantly different from model set A. 

\subsection{Model sets with WRLOF accretion efficiency}
In this section we present the results that we obtain after adopting our model of accretion efficiency in the WRLOF mode of mass transfer. If we consider a spherically symmetric wind from the donor star, the CEMP to VMP ratio is $2.63\%$ when the WRLOF accretion efficiency is proportional to $q^2$ (model set $\mathrm{C}_\mathrm{q}$), and $2.90\%$ in the case of $q$-independent WRLOF (model set C). The corresponding {$\it \gamma2-$}model sets, dubbed $\mathrm{E}_{\mathrm{q}}$ and E, predict a CEMP to VMP ratio of 3.85\% and 4.06\% respectively. 

   \begin{figure*}
   \centering
   \includegraphics[width=0.49\textwidth]{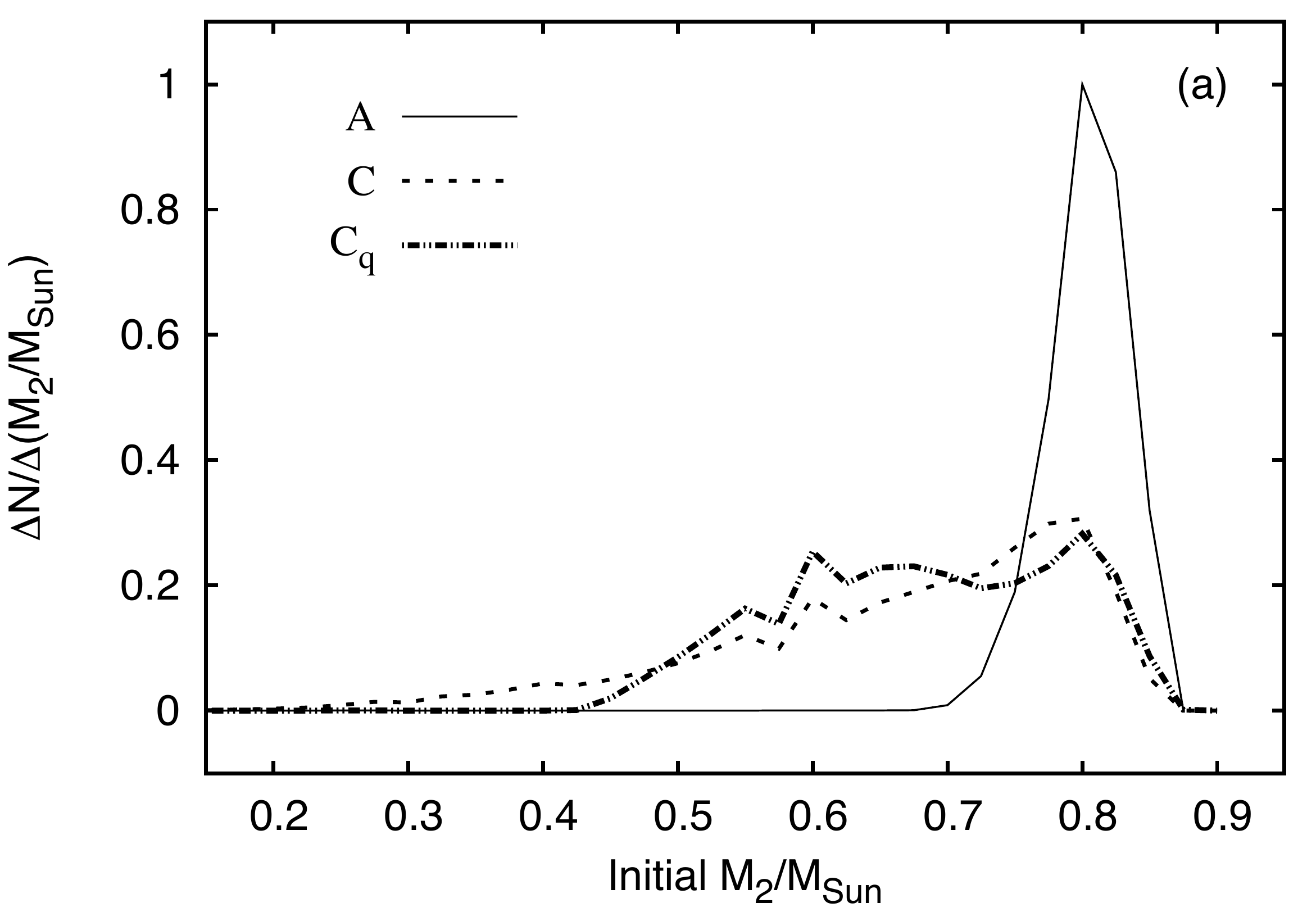}
   \includegraphics[width=0.48\textwidth]{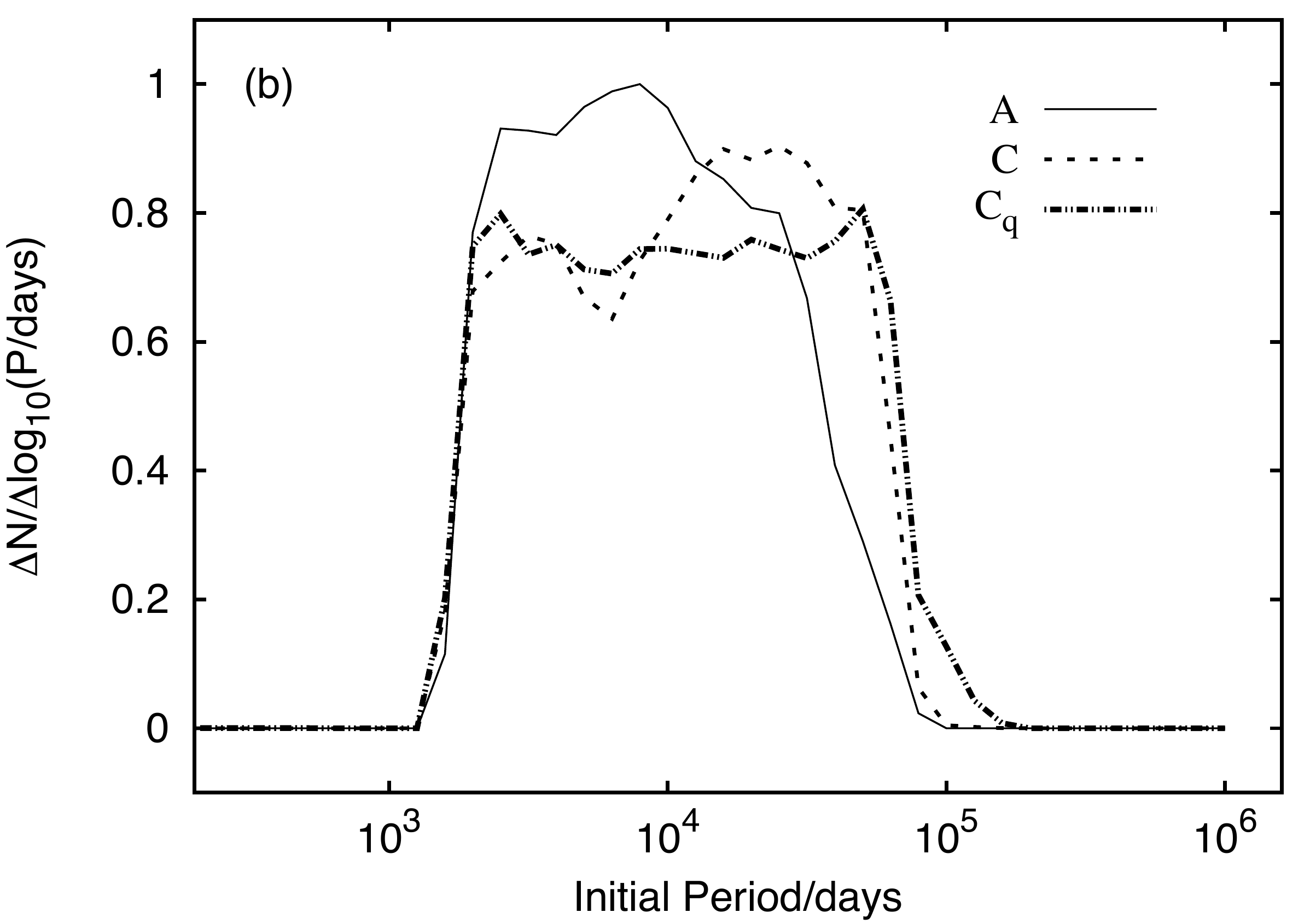}
   \includegraphics[width=0.49\textwidth]{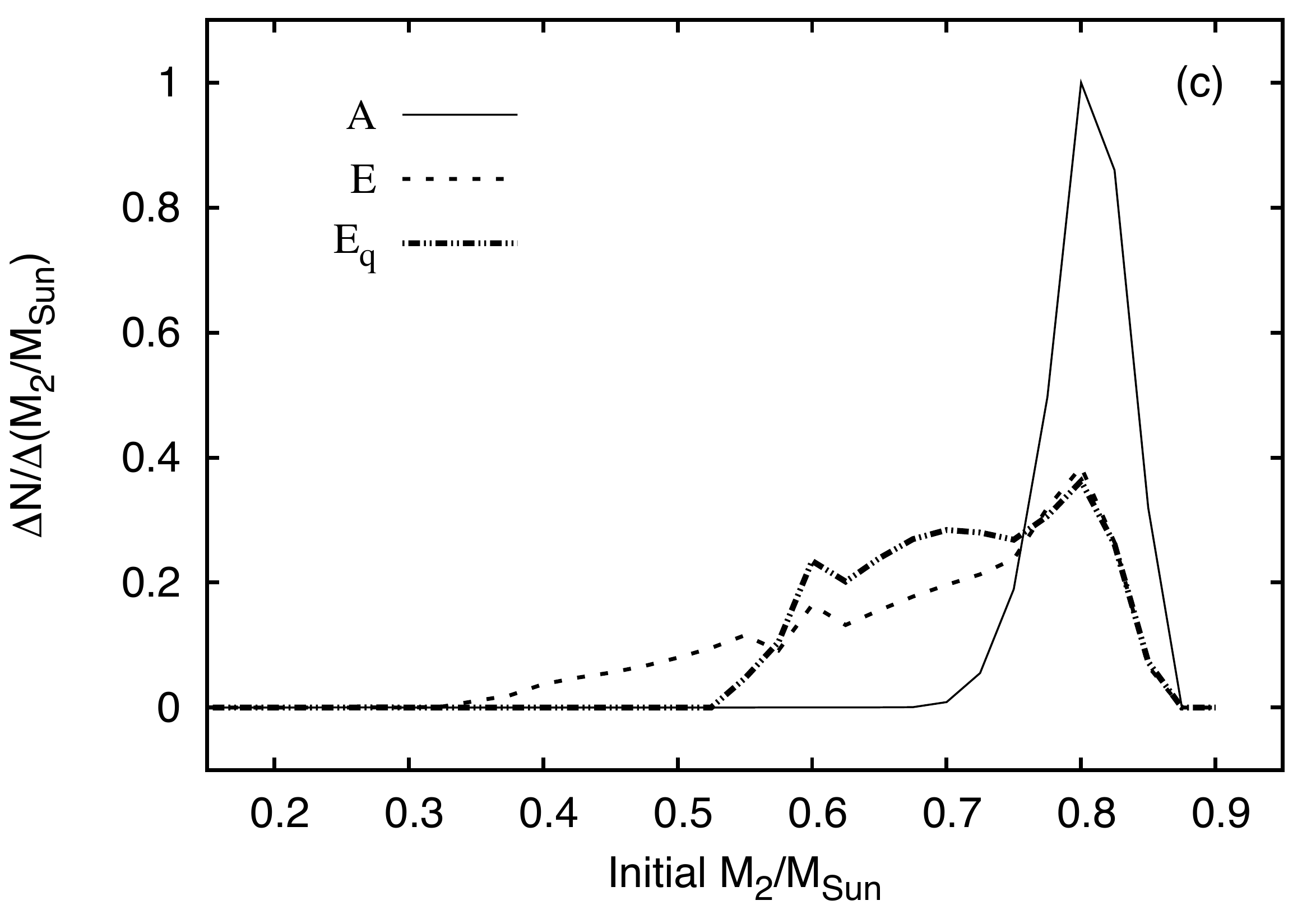}
   \includegraphics[width=0.48\textwidth]{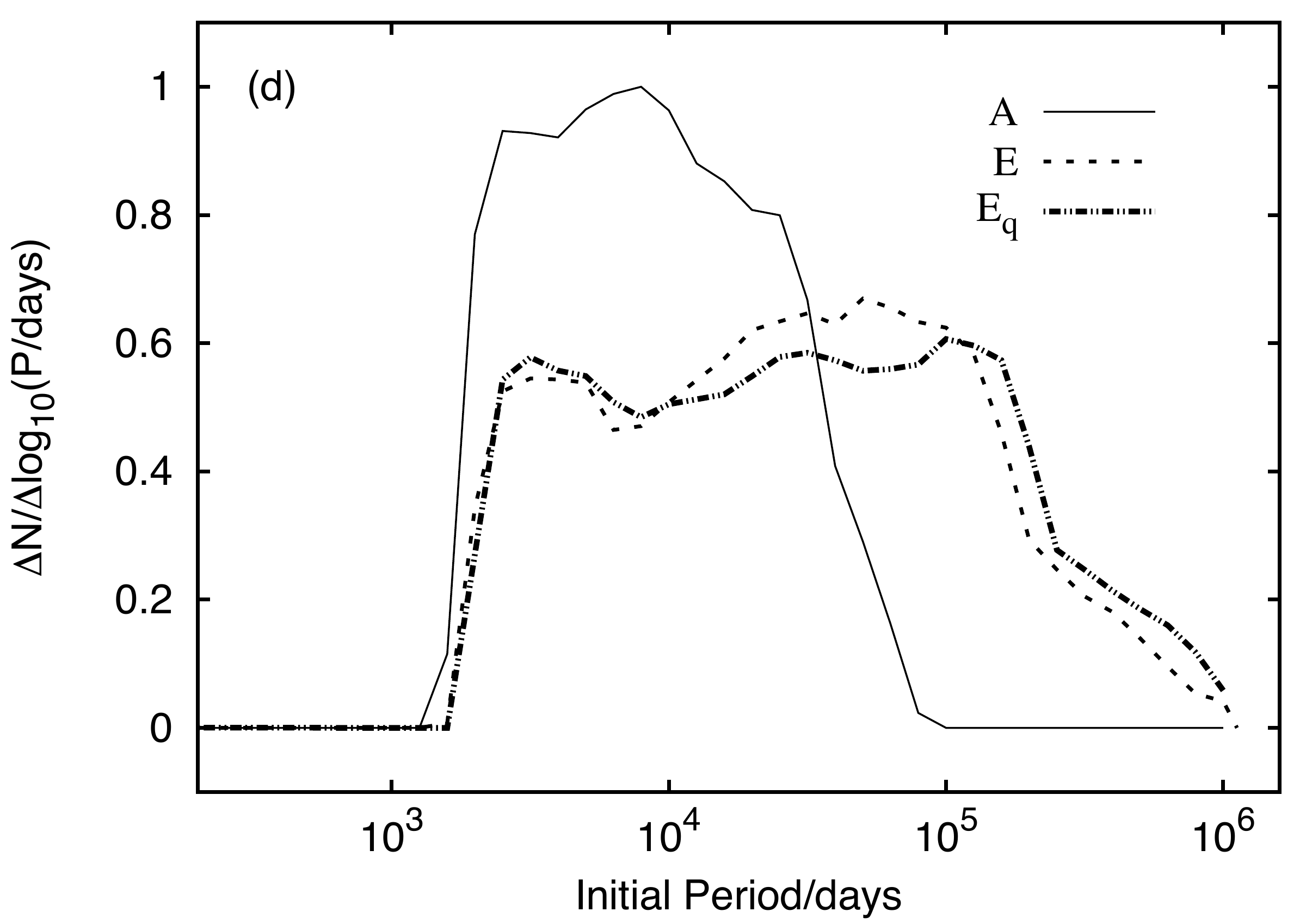}
   \caption{{\bf a)} Distribution of initial secondary mass in binaries that lead to the formation of a CEMP star. {\bf b)} Distribution of initial period of binaries that form CEMP stars. Angular momentum variations are calculated with the spherically simmetric wind assumption (Eq. \ref{eq:jeans}). The $0.1$ dex bins are equally spaced in $M_2/\Msun$ (left panel) or $\log_{10} (P/$days$)$ (right panel). {\bf c)} and {\bf d)} As in (a) and (b), we use $\gamma=2$ in Eq. (\ref{eq:gamma2}) to compute angular momentum loss. Model sets $\mathrm{C}_\mathrm{q}$ and $\mathrm{E}_{\mathrm{q}}$ (dot-dashed lines) are the result of $q$-dependent WRLOF defined by Eq. (\ref{eq:WRLOF-aq}). Model sets C and E (dotted lines) are the result of $q$-independent WRLOF. The solid lines show our default model set A.}
    \label{fig:M2iPi}
    \end{figure*}

\subsubsection{Model sets C and $\mathrm{C}_\mathrm{q}$: CEMP initial parameter space}
The distribution of initial $M_1$ is essentially the same as the corresponding distribution shown in Fig. 4a of I09: it tells us that the majority of CEMP stars are formed by wind accretion from a primary of $1.2-1.5 \,\Msun$. Below this limit the primary does not undergo third dredge up, and therefore no carbon is brought to the surface and transferred to the companion. The highest production of carbon during the TPAGB phase occurs in the mass-range $2-2.5 \,\Msun$, but these stars are much less numerous, according to the IMF of KTG93.

Most notable is the distribution of initial secondary mass shown in Fig. \ref{fig:M2iPi}a: each of the three model sets peak at $0.8 \,\Msun$ but C and $\mathrm{C}_\mathrm{q}$ show a considerable fraction of CEMP stars that come from systems with very low $M_2$ (dotted and dot-dashed line respectively). This low-mass tail is absent if we consider the BHL mode of accretion (model set A, solid line), because with the WRLOF prescription the secondary can accrete mass more efficiently. For example a $1.5\,\Msun$ primary expels around $0.8~\Msun$ of carbon-rich material during its TPAGB phase: hence, accreting 50\% of the lost matter even a star of about $0.4 ~\Msun$ increases in mass sufficiently to evolve to $\log_{10} (g/\mathrm{cm}\,\mathrm{s}^{-2})\le4.0$ when $t >10$ Gyr and be selected as CEMP star. If we observe today such a CEMP star, a large amount of mass we see is from the AGB primary and not the original low-mass companion.

The difference in shape between model sets C and $\mathrm{C}_\mathrm{q}$ results from the mass-ratio dependence of model set $\mathrm{C}_\mathrm{q}$.
In this set, for a given primary mass a low-mass secondary star accretes a smaller amount of material than a more massive secondary star. Therefore the low-mass tail of model set $\mathrm{C}_\mathrm{q}$ does not extend to less than $0.42 \,\Msun$, whereas in set C it extends down to $0.15 \,\Msun$.

Examining the distributions of initial periods in Fig. \ref{fig:M2iPi}b, the effect of WRLOF is to increase the number of wide systems that lead to CEMP star formation. WRLOF accretion is efficient in a wider range of separation than BHL, as illustrated in Fig. \ref{fig:WRLOF-effic}.
Compared to set $\mathrm{C}_\mathrm{q}$, in model set C more CEMP stars come from systems with initial period between $10,\!000$ and $50,\!000$ days. This difference is due to the secondary stars coming from the low-mass tail. To become a CEMP star, a low-mass star needs to accrete material from a relatively massive primary with a large dust formation radius. Therefore such a binary system undergoes the conditions for WRLOF only when the initial separation is relatively wide.

\subsubsection{Model sets E and $\mathrm{E}_{\mathrm{q}}$: CEMP initial parameter space}

The distributions of $M_2$ resulting from model sets E and $\mathrm{E}_{\mathrm{q}}$ (Fig. \ref{fig:M2iPi}c) are qualitatively similar to those obtained from sets  C and $\mathrm{C}_\mathrm{q}$ (Fig. \ref{fig:M2iPi}a). However, the low-mass tails of the distributions of $M_2$ do not extend to as low masses as in the previous sets. A less massive secondary star needs to accrete more mass in order to become a CEMP star, hence it is more likely to form in a system with a relatively high mass primary. Such systems are more likely to fill their Roche lobe and undergo a common-envelope phase in our {\it$\gamma2-$}model sets because of stronger angular momentum loss.
Therefore CEMP stars forming from very low-mass secondaries are relatively rare in the {\it$\gamma2-$}model sets.

The strong angular momentum loss causes the formation of a large fraction of CEMP stars from very wide systems, as can be seen in the initial period distribution (Fig. \ref{fig:M2iPi}d). This effect was already appreciable in the case of BHL wind mass transfer (Fig. \ref{fig:period-y2}) and with the WRLOF mode it is further increased.

The comparison of the initial parameter distributions allows us to compare the effects of the WRLOF and BHL mechanisms, although initial masses and periods are not observable. In the next sections we study the distributions of carbon and nitrogen that are directly comparable to the distributions observed in CEMP stars.

   \begin{figure}
   \includegraphics[width=0.48\textwidth]{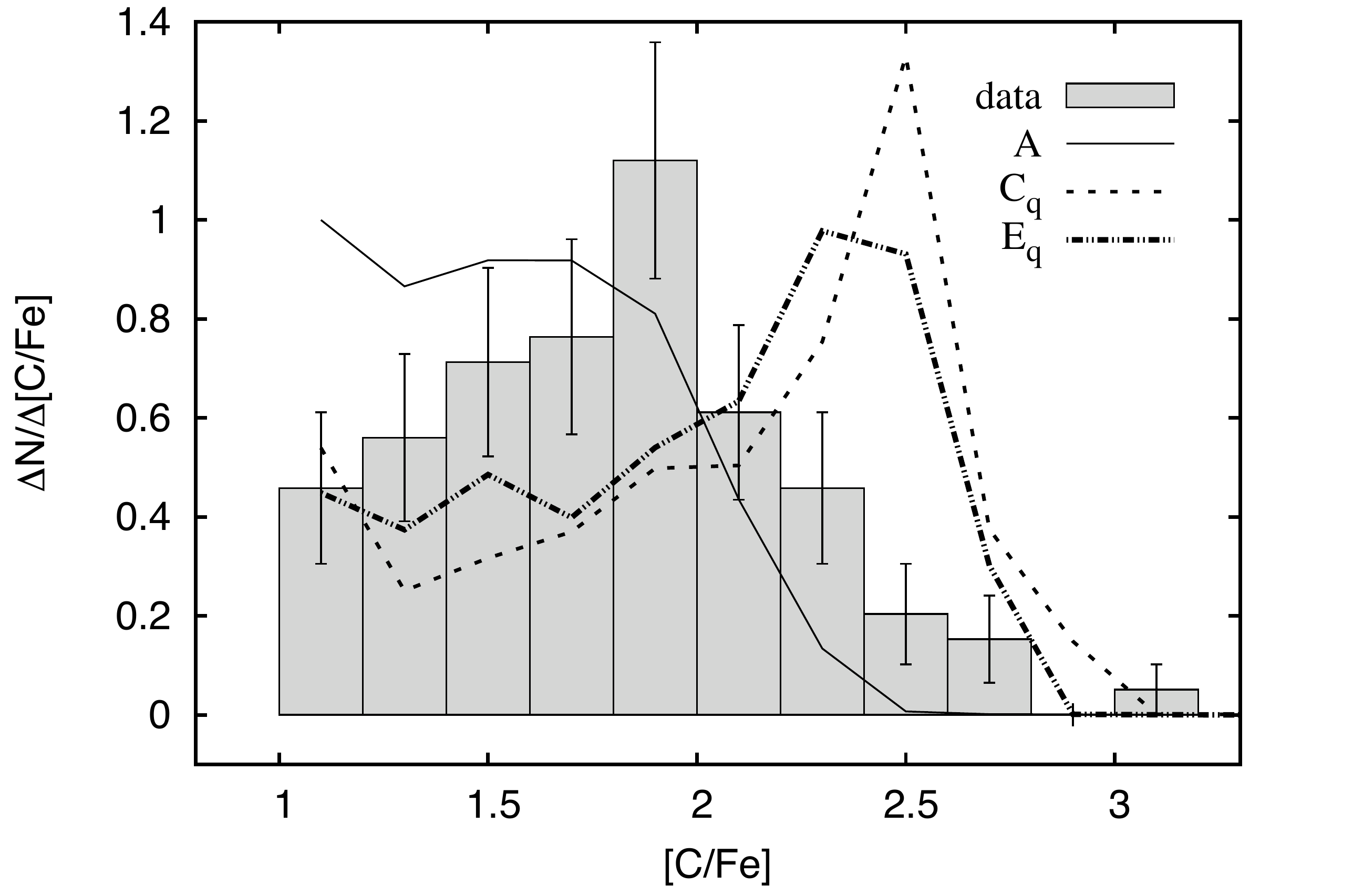}
      \caption{Distribution of [C/Fe] in the CEMP population. The histogram shows the observed distribution in our data sample, with Poisson errors. The solid, dotted and dot-dashed lines are model sets A, $\mathrm{C}_\mathrm{q}$ and $\mathrm{E}_{\mathrm{q}}$ respectively. Model set A is our default, $\mathrm{C}_\mathrm{q}$ and $\mathrm{E}_{\mathrm{q}}$ are the models for $q$-dependent WRLOF. Model set $\mathrm{C}_\mathrm{q}$ assumes a spherically symmetric wind. Model set $\mathrm{E}_{\mathrm{q}}$ is calculated with $\gamma=2$ in Eq. (\ref{eq:gamma2}).}
   \label{fig:abunds}
   \end{figure}
\subsection{Distributions of carbon and nitrogen}
\label{CN}
To compare our models with observations we use a database of VMP stars based on the SAGA observational database compiled by \cite{Suda2008,Suda2011}, combined with data of metal-poor stars from \cite{Frebel2006} and from \cite{Lucatello2006}. We select stars corresponding to our VMP criteria: ($i$) the iron abundance is in the range [Fe/H] $= -2.3\pm0.5$ dex and ($ii$) the star is a giant or a sub-giant, with $\log_{10} (g/\mathrm{cm}\,\mathrm{s}^{-2}) \le 4.0$ (in the database of \citealp{Frebel2006}, stars are catalogued as giants or sub-giants according to their $B-V$ colour). This selection leaves us with a database of 413 VMP stars with measured carbon abundance, of which 100 are CEMP stars. Our database collects all data of VMP and CEMP stars so far available in the literature and is not complete because of the different observational properties and selection effects of the original samples: e.g. data from \cite{Lucatello2006} contains mostly CEMP stars. Therefore the CEMP fraction in our database (approximately $24\%$) is not representative.

Our mass-ratio dependent models $\mathrm{C}_\mathrm{q}$ and $\mathrm{E}_{\mathrm{q}}$ predict abundance distributions of carbon and nitrogen very similar to our $q$-independent models C and E respectively.
Therefore in this section we focus our discussion on the model sets $\mathrm{C}_\mathrm{q}$ and $\mathrm{E}_{\mathrm{q}}$.

None of our model sets is able to accurately reproduce the distribution of carbon in the observed CEMP population, as shown in Fig. \ref{fig:abunds}. 
Our model sets $\mathrm{C}_\mathrm{q}$ and $\mathrm{E}_{\mathrm{q}}$ show qualitatively the same trend as the data. However, the peak around  [C/Fe]$\approx2.5$ corresponds to a carbon abundance $3-4$ times higher than the observations. This might indicate that in our models the accretion efficiency is too high. The fate of the transfered material is to be mixed by the thermohaline process in the envelope of the accretor, composed mainly of hydrogen. However, a star of $0.5 \,\Msun\,$ must accrete $0.3-0.4\,\Msun$ from the donor to be selected as a CEMP star, and this material is only weakly diluted in the envelope. Hence secondary stars with low initial mass become strongly carbon enriched stars. 
In Sect. \ref{WRLOF-mechanism} we estimated the error on the accretion efficiency calculated in the hydrodynamical simulations to be within $50\%$. If we reduce the efficiency of WRLOF by a factor 2 in our model set C$_{\mathrm{q}}$ we remove the stars with initial mass below $0.6\,\Msun$ from the CEMP population and the CEMP/VMP ratio is consequently reduced from $2.63\%$ to $2.53\%$. This also results in a shift of the distribution towards lower values of [C/Fe] but the peak remains around $2.1-2.3$, i.e. higher than the observations. A further decrease in the accretion efficiency in our WRLOF model would presumably result in a better agreement with the observed carbon distribution. However, we should keep in mind that the data in our observed sample may not be representative of the real carbon distribution because our sample is inhomogeneous and incomplete.

\begin{figure}[!]
   \centering
   \includegraphics[width=0.48\textwidth]{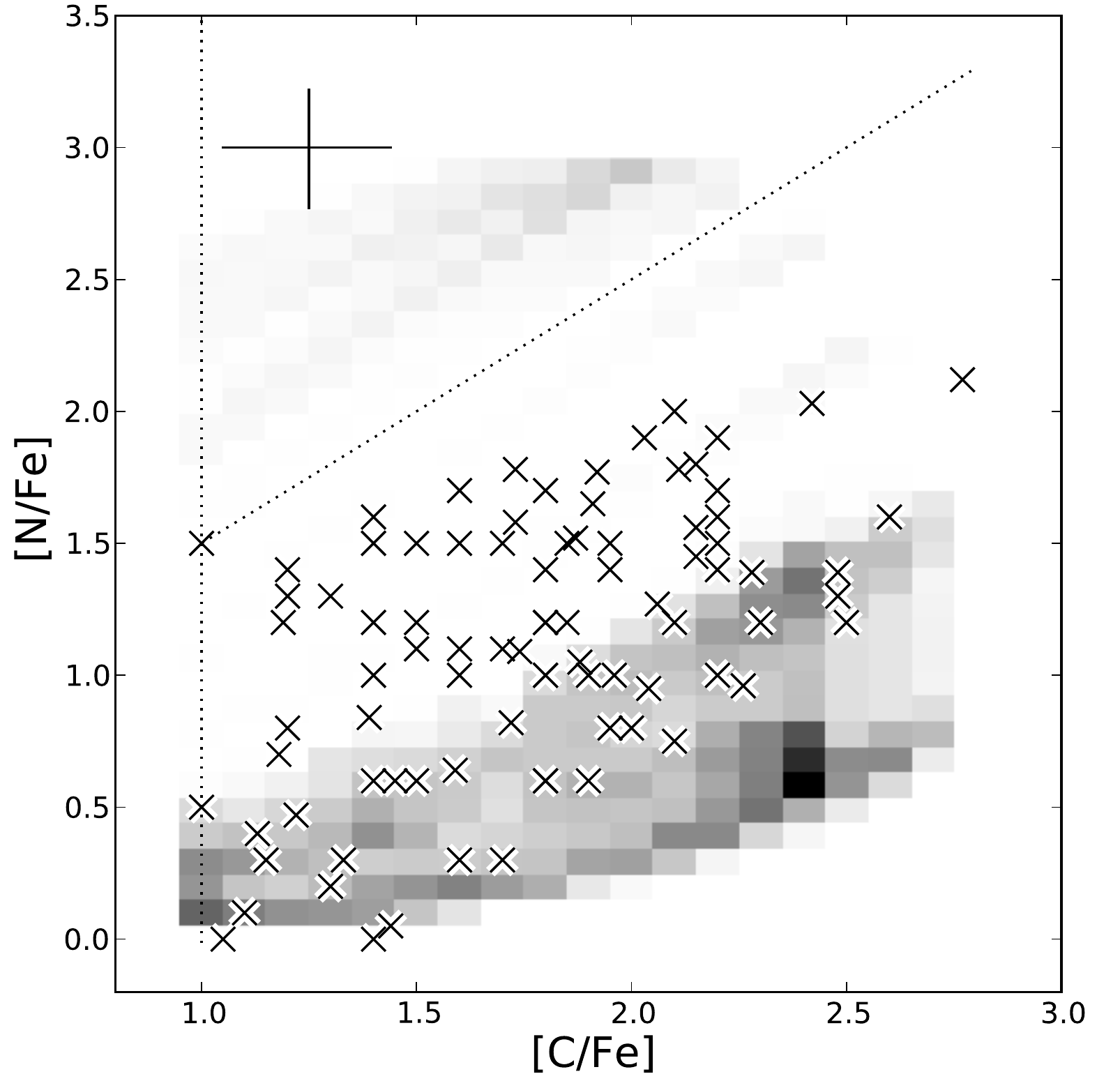}
   \caption{The distribution of [N/Fe] vs [C/Fe] in the CEMP population predicted by $\gamma2$-model set $\mathrm{E}_{\mathrm{q}}$ with $q$-dependent WRLOF. Darker grey indicates greater density of stars. Crosses are data from our selection of SAGA database. Average error bars of the observations are shown in the top-left corner. CEMP stars appear to the right of the vertical dotted line. CNEMP stars enhanced both in carbon and nitrogen appear above the diagonal dotted line.}
         \label{fig:CFeNFe-y2-q2}
   \end{figure}

In Fig. \ref{fig:CFeNFe-y2-q2} we show the distribution of [N/Fe] versus [C/Fe] for {\it$\gamma2-$}model set $\mathrm{E}_{\mathrm{q}}$. Most of our CEMP stars have small nitrogen enhancements and we predict a small number of CNEMP stars, i.e. CEMP stars enhanced in nitrogen ([N/Fe$]\ge 1.0$ and [N/C$]>0.5$). Most observed CEMP stars are in the region between these two groups, where we predict a dearth of systems, and no CNEMP stars are found in our data sample. Our predictions are similar to those of I09 and this is not unexpected, because our updates modify the efficiency of the mass transfer, not the chemical composition of the transferred material. To reproduce the nitrogen enhancement some extra mixing processes are required to burn part of the carbon in the CN cycle and modify the amount of nitrogen brought to the surface of the AGB star (see e.g. \citealp{Nollett2003}).

   \begin{figure}[!ht]
    \includegraphics[width=0.50\textwidth]{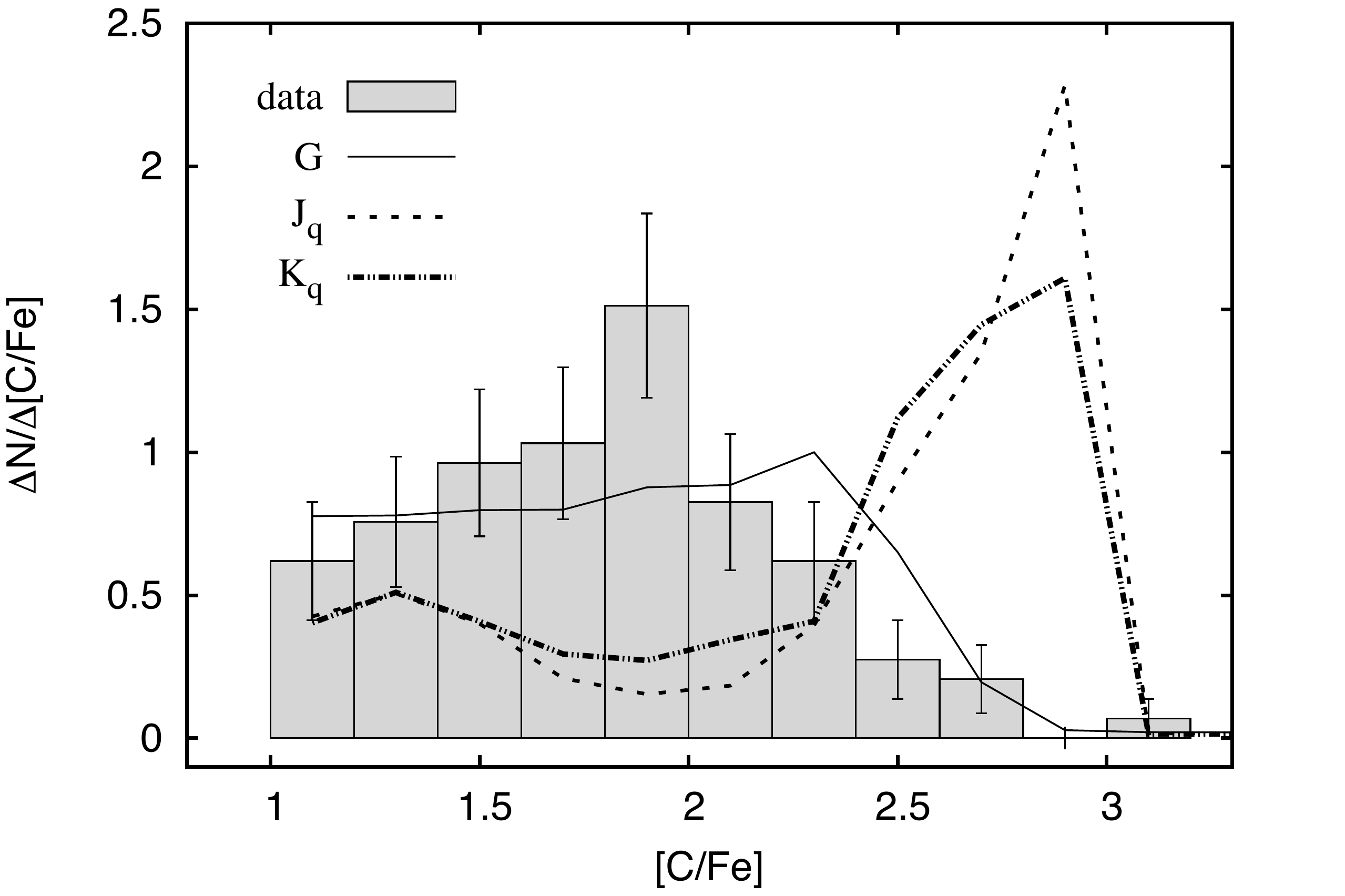}
  \caption{As Fig. \ref{fig:abunds} for models with enhanced third dredge up. Model G (solid line) is calculated with the BHL prescription, $\mathrm{J}_{\mathrm{q}}$ (dotted line) and $\mathrm{K}_{\mathrm{q}}$ (dot-dashed line) are $q$-dependent model sets.}
   \label{fig:abunds-G}
   \end{figure}
\subsection{Model sets with efficient third dredge-up at low mass}
\label{strong3DUP}
In our model populations the fraction of CEMP stars is considerably lower than that observed. Our failure to reproduce the observed distributions of carbon and nitrogen might simply be a consequence of this.
To increase the number of CEMP stars predicted by their standard model set A, I09 choose a set of parameters that reduces the minimum core mass and envelope mass required for third dredge up and increases the amount of material dredged up. With such a set of parameters all TPAGB stars down to initial masses of $0.8 \,\Msun$ undergo efficient third dredge up and I09 find a CEMP/VMP ratio of $9.43\%$, about four times higher than in their standard model A and at the low end of the range of the observations.

The need for efficient dredge up at low core masses and low metallicity has been pointed out by many authors, e.g. \cite{Groenewegen1993, Marigo1996} and \cite{Izzard04L}.
Further justification for these assumptions is provided by recent detailed models that find dredge up for stellar mass down to $0.9\,\Msun$ at $Z=10^{-4}$ (see e.g. \citealp{Stancliffe08}, \citealp{Karakas10}). Indirect indications of efficient dredge up at low mass and metallicity arise from the observed abundance of $s$-elements in metal-poor stars \citep{Axel2007} and from observations of low-mass white dwarfs in the globular cluster M4 (see \citealp{Kalirai2009}).

In our model set G we take the same set of parameters as in model set G of I09 in addition to the equation for the wind velocity proposed by \cite{VW93} and the assumption of a spherically symmetric wind and we obtain a CEMP/VMP ratio of $9.00\%$. The small difference with respect to the result of I09 is due to the different equation for the wind velocity we use to calculate the accretion rate.
When we include the WRLOF prescription in our model G we obtain the same relative increase as in the models without enhanced dredge up. The WRLOF accretion efficiency increases the CEMP star fraction by a factor of about $1.2-1.3$, which further increases to approximately $1.4-1.5$ when we also consider the contribution of efficient angular momentum loss. With the $q$-dependent WRLOF prescription the CEMP/VMP ratio increases to $11.91\%$ in the $ssw$-model set $\mathrm{J}_{\mathrm{q}}$ and to $13.66\%$ in the $\gamma2$-model set $\mathrm{K}_{\mathrm{q}}$. If we assume a binary fraction of $100\%$, these results are within the range of the observations ($9-25\%$).
However, our models predict most of the CEMP stars to have a carbon-enhancement [C/Fe$]>2.5$, much larger than the observations (Fig. \ref{fig:abunds-G}). Because this choice of parameters does not directly affect the amount of nitrogen transported to the stellar surface the discrepancy in nitrogen abundances remains an issue.

   \begin{figure}
   \centering
   \includegraphics[width=0.46\textwidth]{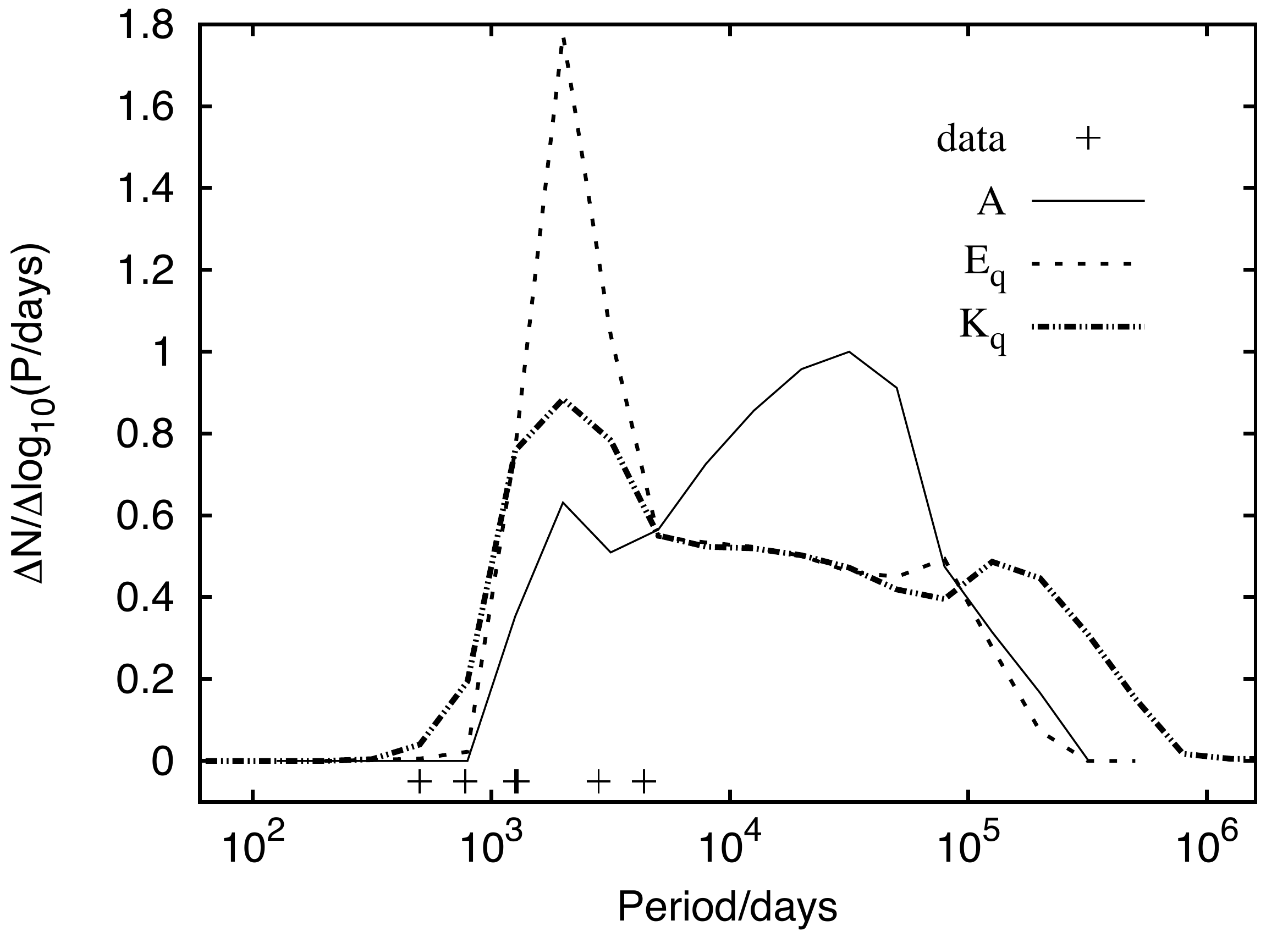}
   \caption{Period distribution in our CEMP population. The 0.2 dex bins are equally spaced in $\log_{10} (P/$days$)$. Solid, dotted and dot-dashed line are model sets A, $\mathrm{E}_{\mathrm{q}}$ and $\mathrm{K}_{\mathrm{q}}$ respectively. Plus signs are the observations from SAGA database. A is the default set. $\mathrm{E}_{\mathrm{q}}$ and $\mathrm{K}_{\mathrm{q}}$ are $\gamma2$-model sets with $q$-dependent WRLOF. Model set $\mathrm{K}_{\mathrm{q}}$ assumes efficient dredge up in stars down to $0.8~\Msun$.}
    \label{fig:P_f}
    \end{figure}
\subsection{Orbital periods}
In Fig. \ref{fig:P_f} we compare the final period distributions from our $\gamma2$-model sets $\mathrm{E}_\mathrm{q}$ and $\mathrm{K}_{\mathrm{q}}$ to our default model A and to the data from our selection of the SAGA database. 
Model set $\mathrm{E}_\mathrm{q}$ predicts most CEMP stars at orbital periods around $2000$ days, whereas default model set A predicts mostly wider systems and peaks around $40,\!000$ days. Model set $\mathrm{K}_\mathrm{q}$, which allows efficient third dredge up in low-mass stars, predicts CEMP stars in a range of orbital periods wider than model set $\mathrm{E}_{\mathrm{q}}$ and is able to reproduce the shortest-period CEMP stars of our sample. A similar result is obtained with model sets G and $\mathrm{J}_{\mathrm{q}}$.

Even though many CEMP stars are known to be binaries, only a few have known orbital periods because long periods are difficult to measure. In our observed sample $6$ out of $100$ CEMP stars show a period below $4400$ days (the longest-period measured in a CEMP star so far), i.e. a fraction of $0.06$ if we assume that all the other observed CEMP stars are in binary systems with a longer period. In reality this is a lower limit because only a fraction of the observed systems in our heterogeneous sample have been investigated to determine the period. Model sets A, $\mathrm{E}_\mathrm{q}$ and $\mathrm{K}_\mathrm{q}$ predict a fraction of CEMP stars with period shorter than $4400$ days of $0.2\,$, $0.48$ and $0.36$ respectively (values in the same range are found with all the other model sets). This exceeds the observational lower limit, and therefore we are not able to rule out any of our models.

\section{Discussion}
\label{discussion}
This paper presents the first study in which the WRLOF mode of mass transfer is taken into account in binary population synthesis: the purpose of this work is to investigate how the implementation of a more realistic description of the wind mass transfer process affects a population of CEMP stars. 

WRLOF has two effects on the evolution of binary systems: it increases both the accretion efficiency of the mass transfer and the angular momentum taken away by the material lost in the wind. 
The enhanced accretion efficiency produces an increase in the CEMP to VMP number ratio by a factor $1.2-1.3$. It has strong effects on the initial secondary mass distribution, which gains low-mass secondary stars, and on the carbon distribution, which peaks at large [C/Fe].
Our prescription of efficient angular momentum loss increases the CEMP to VMP ratio by a factor $1.2-1.5$ and has strong effects on the initial and final period distributions of CEMP stars. Initially wide systems shrink enough to interact and transfer mass to the secondary star. Moreover, a higher number of systems are predicted at relatively short final periods.
The combined contribution of these two effects widens the initial range of systems that become CEMP stars towards longer periods and lower secondary masses. As a result the CEMP to VMP ratio predicted by our models increases by a factor $1.4-1.8$ and we find a distribution of carbon abundances more similar to observations than in previous studies.

The reliability of our model of WRLOF is limited by the small number of hydrodynamical simulations of wind accretion in binary systems. 
In particular, the efficiency of angular momentum loss by the stellar wind is poorly constrained by currently available models. We have assumed two different, very simple models: a spherically symmetric wind that does not interact with the binary system (equivalent to $\gamma=q\boldsymbol{\le 1}$ in eq.~\ref{eq:gamma2}), and a strongly modified wind that carries away twice the average specific angular momentum of the orbit ($\gamma=2$). 
These assumptions may respresent very wide systems and relatively close systems, respectively, because the specific angular momentum carried away by the wind is likely to decrease with increasing ratio of wind velocity over orbital velocity \citep{Jahanara2005}.  The dependence of $\gamma$ on orbital period and mass ratio cannot be quantified further at present, and our simple assumptions can only bracket the real situation.
To improve upon our model of the enhanced accretion rate and of the efficiency of angular momentum loss more hydrodynamical simulations are needed, and for a wider range of initial masses and separations.
Considering the uncertainties related to the accretion rate, we tested how strongly the CEMP to VMP fraction depends on the accretion efficiency in the WRLOF regime. The dependence is found to be weak: if the accretion efficiency is reduced by a factor 2, the CEMP/VMP ratio decreases from $2.90\%$ to $2.77\%$ in model set C and from $2.63\%$ to $2.53\%$ in model set C$_{\mathrm{q}}$.

Another uncertainty is related to the condensation temperature used to determine the dust formation radius of AGB stars. A variation in $T_{\mathrm{cond}}$ affects the range of separations over which binary systems undergo WRLOF. We tested this by varying $T_{\mathrm{cond}}$ by $10\%$ around the assumed value of $1500\,$K. A lower condensation temperature ($T_{\mathrm{cond}}=1350\,$K) does not affect the CEMP/VMP ratio while the initial and final period distributions include more wide systems. If $T_{\mathrm{cond}}$ is $1650\,$K the CEMP/VMP in model C is decreased from $2.90\%$ to $2.50\%$ because the number of CEMP stars with initial period longer than $30,\!000$ days is almost zero and initially close systems, with period shorter than 1000 days, enter into a common envelope which prevents mass accretion.
Our results are also affected by uncertainties related to the AGB phase, such as the amount of carbon and nitrogen dredged up to the surface and the wind acceleration mechanism, to the mixing processes that occur in the primary and the secondary star and to the IMF at low metallicity. 

We have assumed that all stars are formed in binary systems with semi-major axes $3 < a/\Rsun < 10^5$. The observed fraction of binary systems with this range of orbits in the solar neighbourhood is smaller, however.
\cite{DuquennoyMayor1991} find that in the solar neighbourhood about $60\%$ of solar-like stars have a binary companion. The study by \cite{Kouwenhoven2007} of the young stellar association Scorpius OB2 shows a binary fraction of at least $70\%$ ($3\sigma$ confidence) and probably close to unity among stars of spectral type A and B in the range $5 \Rsun \lesssim a \lesssim 5\times 10^6 \Rsun$. 
With the assumed flat distribution in $\log_{10}a$, the range of separations in our grid corresponds to about $75\%$ of the range analysed in Scorpius OB2. 
This implies that our calculated CEMP/VMP ratios should be reduced by a factor $\lesssim 0.75$, although we note that the fraction and distribution of binary systems among metal-poor halo stars is not well constrained and may be different than in the solar neighbourhood.

WRLOF can influence a variety of astrophysical phenomena outside the context of this paper \citep{Shazrene07}. For instance, WRLOF could likely play a role in the wind-accretion scenario which is generally accepted to explain the formation of the barium and CH stars observed at relatively higher metallicity. Also, in the case of formation of planetary nebulae in binary systems WRLOF widens the range of separations in which the presence of a companion star distorts the shape of the ejected material. Finally, in a symbiotic binary the white dwarf could accrete enough mass from an AGB companion star to trigger nova outbursts or the explosion of a Type Ia supernova. It would be interesting to study the effects on these phenomena of our implementation of WRLOF.

\section{Summary and conclusions}
\label{conclusions}

We summarise here our analysis.
\begin{enumerate}
\item[$i$)]We have updated our population synthesis code by improving algorithms to calculate the angular momentum variations and wind velocity. These changes result in a modest change in the overall population of carbon-enhanced metal-poor (CEMP) stars. 
\item[$ii$)]We propose a model for the wind Roche-lobe overflow (WRLOF) mode of mass transfer. In this model the accretion efficiency of mass transfer is based on the results of hydrodynamical simulations of \cite{Shazrene10} and only depends on the radius of the donor star and on the separation of the binary system. We also introduce a dependency on the mass ratio of the two stars that scales as $q^2$. We consider the effect of WRLOF on the orbital angular momentum lost by the binary systems with a simple prescription in which the wind carries away twice the average specific angular momentum of the orbit.
\item[$iii$)]The WRLOF accretion efficiency causes an increase by a factor $1.2-1.3\,$ of the number ratio of CEMP stars to VMP stars. The prescription for efficient angular momentum loss increases the CEMP/VMP ratio by a factor $1.2-1.5$. When both effects are taken into account we predict a CEMP/VMP ratio up to $4\%$ if we assume a binary fraction of unity.
\item[$iv$)]Our model with efficient third dredge-up in stars of mass down to $0.8\,\Msun$ predicts a CEMP/VMP ratio of $9\%$, confirming the results of \cite{Izzard09}. When we combine efficient dredge up with the WRLOF model we find a similar relative increase in the CEMP/VMP ratio as predicted in the models with standard dredge up and we predict a CEMP/VMP ratio in the range $11-14\%$ (again for a binary fraction of unity).
\end{enumerate}

In conclusion, our model of WRLOF widens the range of binary systems which form CEMP stars and as a result increases the predicted CEMP fraction and modifies the distribution of carbon abundance and period of the CEMP population. Assuming in addition efficient dredge up in low-mass stars we obtain CEMP/VMP ratios in the range of the observations. Other physical processes need to be considered to reproduce the distribution of carbon and nitrogen observed in CEMP stars.

\begin{acknowledgements}
We thank the referee C. A. Tout for his detailed comments that have helped to improve the clarity of this paper. CA thanks the NWO for funding support. SdM acknowledges NASA Hubble Fellowship grant HST-HF- 51270.01-A awarded by STScI, operated by AURA for NASA, contract NAS 5-26555.
\end{acknowledgements}

\end{document}